\documentclass[%
 reprint,
nofootinbib,
 amsmath,amssymb,
 aps,
]{revtex4-1}

\usepackage{amsmath}
\usepackage{amssymb}
\usepackage[usenames, dvipsnames]{color}
\usepackage{graphicx}% Include figure files
\usepackage{dcolumn}% Align table columns on decimal point
\usepackage{bm}% bold math
\usepackage{color}
\usepackage[section]{placeins}
\usepackage{hyperref}
\allowdisplaybreaks

\usepackage{listings}
\usepackage{color}

\definecolor{dkgreen}{rgb}{0,0.6,0}
\definecolor{gray}{rgb}{0.5,0.5,0.5}
\definecolor{mauve}{rgb}{0.58,0,0.82}

\newcommand\mycdots{\hbox to 0.9em{$\cdot$ \hss $\cdot$ \hss $\cdot$}}

\begin{document}

\preprint{APS/123-QED}

\title{Particle-Hole Mirror Symmetries around the Half-Filled Shell:\\
The Quantum Numbers and Algebraic Structure of Composite Fermions}% Force line breaks with \\

  \author{W. C. Haxton}
% \altaffiliation[Also at ]{Department of Physics,  University of California, Berkeley.}%Lines break automatically or can be forced with \\
\email{haxton@berkeley.edu}
\affiliation{Department of Physics,  University of California, and \\ Nuclear Science Division, Lawrence Berkeley National Laboratory, Berkeley, CA}
\author{Daniel J. Haxton}
 \email{danhax@gmail.com}
 \affiliation{Department of Physics, University of California, Berkeley, CA  and KLA-Tencor, Fremont, CA}
 \author{Byungmin Kang}
 \email{bkang119@berkeley.edu}
 \affiliation{Department of Physics,  University of California, and \\ Materials Sciences Division, Lawrence Berkeley National Laboratory, Berkeley, CA}


\date{\today}% It is always \today, today,

\begin{abstract}
Composite fermions (CFs) of the fractional quantum Hall effect (FQHE) are described as spherical products of electron and vortex spinors, 
built from underlying $L={1 \over 2}$ ladder operators aligned so that the spinor angular momenta, $L_e$ and $L_v$,
are maximal.  We identify
the CF's quantum numbers as the angular momentum $L$ in $(L_eL_v)L$, its magnetic projection $m_L$,
the electron number $N$ ($L_v={N-1 \over 2}$), and magnetic
$\nu$-spin,  $m_\nu=L_e-L_v$.  Translationally invariant FQHE states are formed by fully filling $p$ subshells with 
their respective CFs,  in order of ascending $L$ for fixed $L_e$ and $L_v$, beginning with the lowest allowed value, $L=|m_\nu|$.
CF subshells are contained entirely within the first Landau level (FLL).  Alternatively, we provide an equivalent hierarchical wave function
in which the underlying objects are vortices with $L_v={p \over 2}$,
correlated pairwise via $\vec{L}_{v_i} \cdot \vec{L}_{v_j}$.

We show that CFs can be written as a valence operator carrying the angular momentum quantum numbers $L,m_\nu,m_L$
acting on a scalar half-filled intrinsic state.  This scalar state  serves as the vacuum for the valence electron 
($\boldsymbol{b}^\dagger,\boldsymbol{\tilde{b}}$) and vortex ($\boldsymbol{v}^\dagger,\boldsymbol{\tilde{v}}$)
ladder operators that create FQHE states.

With respect to this vacuum, FQHE states can be grouped into $\nu$-spin multiplets mirror symmetric around $m_\nu$=0,
in which $N$ is held constant.  $m_\nu \ne 0$ states have a net electron particle or hole number.
Particle-hole conjugation with respect to this vacuum is identified as the mirror symmetry relating FQHE states of the same $N$ but distinct fillings 
$\nu = {p \over 2p+1}$ and $\bar{\nu}= {p \over 2p-1}$, e.g.,  $2/5 \leftrightarrow 2/3$.

Alternatively, mirror symmetric $\nu$-spin multiplets can be constructed in which the magnetic field strength is held fixed: the valence
states are electron particle-vortex hole excitations relative to the half-filled vacuum ($m_\nu>0$)
and their mirror conjugates ($m_\nu<0$).  Multiplet members are linked by the $\nu$-spin raising/lowering operators,
$\hat{S}^\nu_\pm$.   Particle-hole (PH) symmetry -- relating the $N$-particle FQHE state $\Psi$ of filling
$\nu={p \over 2p+1}$ to the $\bar{N}$-particle state $\bar{\Psi}$ of filling $\bar{\nu}={p+1 \over 2p+1}$, e.g.,  $2/5 \leftrightarrow 3/5$ --
is shown to be equivalent to electron-vortex exchange, $\boldsymbol{b}^\dagger \leftrightarrow \boldsymbol{v}^\dagger$
and  $\boldsymbol{\tilde{b}} \leftrightarrow \boldsymbol{\tilde{v}}$.   The $N$-particle states $\Psi$ and $\hat{S}_- \bar{\Psi}$ are
connected by this mirror symmetry.  

In this construction $\bar{N}-N$ CFs of the state $\bar{\Psi}$
occupy an extra zero-mode subshell that is annihilated by $\hat{S}^\nu_-$.  We link this structure, familiar from supersymmetric quantum mechanics, to the CF Pauli Hamiltonian, which 
we show is isospectral, quadratic in the $\nu$-spin raising and lowering
operators $\hat{S}^\nu_\pm$, and four-fold degenerate in $\Psi$, $\hat{S}^\nu_- \Psi$, $\bar{\Psi}$, and
$\hat{S}^\nu_- \bar{\Psi}$.  On linearization, it takes a Dirac form 
similar to that found in the integer quantum Hall effect (IQHE).

\iffalse
\begin{description}
\item[Usage]
Secondary publications and information retrieval purposes.
\item[PACS numbers]
May be entered using the \verb+\pacs{#1}+ command.
\item[Structure]
You may use the \texttt{description} environment to structure your abstract;
use the optional argument of the \verb+\item+ command to give the category of each item. 
\end{description}
\fi
\end{abstract}

\pacs{Valid PACS appear here}% PACS, the Physics and Astronomy
\maketitle

%\tableofcontents

% As a matter of principle the effective theory should be matched to observables.
% In 
% The harmonic oscillator basis has the advantages of center of mass separability.

\section{Introduction}
 Laughlin's wave function \cite{Laughlin} 
for the fractional quantum Hall effect \cite{Tsui}, describing the $\nu=1/3$ state, was later
extended by Jain \cite{Jain} to the series of fillings
 $\nu = p/(2p+1)$, $p=1,2, \dots$.  The two constructions differ.  Laughlin's was carried out in
 the FLL, and employed a
 variational argument to  constrain the two-electron correlation function.
  In contrast, Jain's construction made no reference to the electron-electron interaction,
 but instead employed closed-shell states from the IQHE.  He postulated a noninteracting form for FQHE wave functions similar to
 that of the IQHE, borrowing from that problem the needed operator structure.
 As the resulting wave functions span several LLs, numerical projection was required to produce FLL trial wave functions.
 
 Certain conceptual issues presented by Jain's construction were addressed by Ginocchio
 and Haxton (GH) \cite{GH}, who argued that Laughlin's variational argument could be extended
 to successively larger groups of electrons, producing a generalized set of FLL closed shell
 operators  with Laughlin's $\nu=1/3$ operator being the first member of the series.  The GH operators have
 an $\ell s$ structure associated with placing an electron in the plane ($\ell$) and destruction of magnetic
 flux ($s$) (e.g., removing factors of $u(i) \cdot u(j)$ that keep electrons separated).  GH applied their operators to the $\nu=1/2$ bosonic state
 to generate analytic FLL wave functions for both hierarchy states ($\nu=1/3, 2/5, \dots$) and their
 conjugates ($\bar{\nu}=1,2/3, 3/5, \dots$).  When numerically evaluated, these wave functions were
 found to be identical to those of Jain, to the accuracy to which the latter had been evaluated:
 this is a consequence of the SU(2) algebra both procedures impose on wave functions, as we later discuss.
 We refer to wave functions constructed with numerical projection as Jain wave functions,
 and those  constructed analytically with FLL GH operators as the Jain/GH wave functions.
 
 Jain's interpretation of his wave function is the basis for one of the field's most important paradigms, 
 the existence of CFs \cite{JainCF}: the open-shell, strongly interacting electron states at $\nu=1/3, 2/5, \cdots$
 can be represented as closed-shell noninteracting CF states.
 To our knowledge, however, no one has yet to write the Jain wave function exactly in such a form,
 generating some distress \cite{dyakonov}.
 
 Recently a single-determinant CF form of FQHE wave functions was obtained
 by applying the GH scalar operators to the half-filled shell in a slightly different way.
 The GH$^2$ construction \cite{GH2} remains consistent with the essentials of Jain's picture -- specifically
 1) CFs behaving as if they were IQHE electrons within a reduced B-field \cite{Note} and 2) the absorption of two units of
 magnetic flux into CF internal wave functions with the addition of each new electron.  GH$^2$ CFs 
 are a spherical product of an electron spinor of angular momentum $L_e$ and a vortex
 spinor with $L_v = {N-1 \over2}$, where $N$ is the electron number.   These spinors are themselves generated
 from the aligned coupling of $n_e \equiv 2L_e$ and $n_v \equiv 2L_v$ elementary $L$=${1 \over 2}$ spherical
 creation operators, respectively.  Thus the CF for electron 1 is
 \[\left[ [\boldsymbol{b}^{\dagger \, n_e}_1]^{L_e} \otimes [\boldsymbol{b}^{\dagger}_2 \cdots \boldsymbol{b}^{\dagger}_N]^{L_v} \right]^L_{m_L} .\]
 For fixed $L_e$ and $L_v$, one can form families of CFs
 indexed by the CF angular 
 momentum $L$,  $(L_e L_v)L$.  This CF structure allows one to form many-electron states of total 
 angular momentum $L^T=0$ simply by filling all 2$L$+1 distinct magnetic substates $m_L$ in a given CF subshell.
 GH$^2$ FQHE states correspond to filling the lowest $p$ such subshells,
 beginning with the energetically favored anti-aligned coupling that produces the lowest permissible $L=|L_e-L_v|$, and incrementing.
 The states have fillings of $v={p \over 2p+1}$ or ${p \over 2p-1}$, $p=1,2,3, \dots$, depending on the sign of $L_e-L_v$.
 Such $L^T=0$ states are both translationally invariant and homogeneous (uniform one-body density over the sphere),
 properties required of FQHE states.
 
 This description of CFs differs from that most frequently seen in the literature, where CFs are argued to 
 be an electron coupled to two units of magnetic flux.  In our view this
 reflects confusion between the recursion relation for FQHE wave functions and the quasiparticles (the CFs) 
 from which simple noninteracting, closed-shell many-electron wave functions are formed.
 For example, in the case $\nu={1 \over 3}$ state, two units of flux are indeed absorbed
 into the intrinsic wave functions of the CFs when one adds an electron to an $N$-electron wave function, to
 form a new $\nu={1 \over3}$ wave function of $N$+1 electrons.  However only one unit is 
 associated with the $N$+1 CF being added: the second is absorbed, one quantum each, into the vortices of the 
 $N$ pre-existing CFs, increasing $L_v$ from ${N-1 \over 2}$ to ${N \over 2}$.  The GH$^2$ construction is consistent with the fermionic character of CFs.
 
 In this paper we present several new results connected with the quantum numbers of CFs, their algebraic substructure,
 the symmetries that emerge from understanding CF quantum numbers and properties, and the CF effective Hamiltonian.   We show that 
 CFs carry a second magnetic quantum number associated with $L$, $\nu$-spin or $m_\nu=L_e-L_v={n_e -n_v \over 2}$,
 which allows one to organize FQHE states into multiplets that exhibit mirror symmetry with respect to the half-filled shell.
 $\nu$-spin is so named because it governs symmetries that relate
 states of different filling $\nu$, and because of the analogy to QCD's isospin, $m_\tau$,
 which measures the difference in the up and down quark number.  The mirror symmetries associated with $\nu$-spin
 are strikingly similar to the isospin mirror symmetries that connect the CFs of QCD, the $m_\tau= \pm {1 \over 2}$ 
 proton and neutron, as well as the
 richer isospin multiplets that arise in many-nucleon systems.
 
 Because some of the following sections are somewhat algebraic, we summarize here the motivation and main results
 of each section, to help guide the reader.
 
 Section II describes the IQHE in spherical geometry, which we include because the algebraic structure of this
 problem resembles that of the FQHE, once the mapping to CFs is done.
 The most commonly used set of elementary basis functions for the IQHE are those
 introduced by Haldane \cite{Haldane}, $u$, $v$, ${u}^*$, and ${v}^*$.  We point out that these are the four magnetic components
 of a spinor carrying two indices
 \[  \mathcal{D}^{L={1 \over 2}}_{m_S={\pm {1 \over 2}} \, m_L={\pm {1 \over 2}}} (0,\theta, -\phi) ~~ \leftrightarrow~~ \boldsymbol{c}^\dagger_{m_S \, m_L}|0 \rangle   \tag{a} \label{tIQHE} \]
 associated with creating single quanta with $L={1 \over 2}$.
 Here $\theta$ and $\phi$ are the usual angles for the unit sphere, and $\mathcal{D}$ is the rotation matrix.  These solutions
 correspond to an elementary monopole (positive or negative) at the
 sphere's center.  The general
 solution for arbitrary monopole strength can be built up from ${2L}$ elementary spinors as an aligned product, $\mathcal{D}^{L}_{m_S \, m_L} (0,\theta, -\phi)$.
 This structure - a total angular momentum $L$ and two magnetic quantum numbers $m_S$ and $m_L$ associated with two distinct, commuting
 SU(2) algebras - is identical to that we later develop for the FQHE.    The second magnetic quantum number $m_S$ is the eigenvalue of an operator $\hat{S}_0 \equiv \hat{S}_z$ introduced by Haldane \cite{Haldane}: $2m_S$ is the number of monopoles at the center of the sphere.
 The associated raising and
 lowering operators $\hat{S}_\pm$, discussed by Greiter \cite{Greiter}, can be used to move wave functions in magnetic field space.
 They provide a simple representation of the IQHE Hamiltonian
 \[\hat{H} = {\hbar \omega \over 2 \, m_S}~ {1 \over 2} \left[ \hat{S}_+\hat{S}_- +\hat{S}_-\hat{S}_+ \right]   \tag{b} \label{tIQHEH}.\]
 
 In Section III we describe the CF representation of FQHE wave functions on the sphere derived in \cite{GH2}.
 Electrons in a neutralizing background minimize their energy by 
 keeping their distance from one another: a CF is an electron dressed by a translationally invariant intrinsic 
 wave function that builds in this separation through factors of $u_i \cdot u_j$, where $i$ is the electron label and $j \ne i$.  The dressed electron thus carries
 a somewhat unusual void: the void guarantees a diminished electron density in the immediate vicinity of the electron to 
 which it is attached, encasing that electron in a positively charged cocoon, but also affects correlations with distant electrons.   For $\nu={1 \over 3}$ the void includes a $u_i \cdot u_j$ pair for every $j$, and all $N$ CFs are equivalent, filling a single subshell ($p=1$) with $L={N-1 \over 2}$.
 For $\nu>{1 \over 3}$, the $u_i \cdot u_j$ correlation cannot be maintained for all $j \ne i$:  we show this leads to a
 subshell structure within FLL, with each subshell associated with a distinct CF.    $L^T=0$
 many-CF states can still be constructed by fully filling multiple CF subshells
 ($p>1$).  The subshells are indexed by $\mathcal{I}$, $1 \le \mathcal{I} \le p$, with the
 CFs in shell $\mathcal{I}$ maintaining favorable scalar-pair correlations with
 $N-\mathcal{I}$ neighbors, only.  The $\mathcal{I}-1$ missing antisymmetric pairs become symmetric pairs, with each pair generating an additional
 unit of angular momentum.  These symmetric pairs produce an ascending angular momentum tower $L(\mathcal{I})=|m_\nu|+\mathcal{I}-1$.   
 Thus the CFs occupying different subshells are distinguished by the $L$ labels they carry.
 
 Section III also presents a new, hierarchical form
 of the GH$^2$ wave function in which the underlying objects are vortices of $p$ electrons
 with spin ${p \over 2}$, $[u_1 \otimes \cdots \otimes u_p]^{p \over 2}$, attached to
 a scalar intrinsic wave function that corresponds to a closed shell of $p$ CFs.  The ${N \over p}$ vortices are then kept apart in the
 simplest possible way, pairwise spin-spin correlations.  The Laughlin $\nu={1 \over 3}$ wave function is the $p=1$ limit where the CF and hierarchical
 wave functions are in fact identical in form.  Thus the Laughlin wave function can be considered the ``seed" for either the CF
 or hierarchical series of wave functions.
 
 Section IV develops the algebraic structure of CFs.  It begins by identifying the full set of CF quantum numbers, $N$, $L$, $m_L$, and
 $m_\nu= \ell-s$, introducing the $\nu$-spin quantum number that governs mirror symmetries around $\nu={1 \over 2}$
 ($m_\nu=0$).   The use of $m_\nu$ allows us to organize FQHE wave functions of different fillings but the same electron number $N$ into
 angular momentum multiplets, symmetric around $m_\nu$=0.    When the associated CFs are written with respect
 to a vacuum state $|0_N \rangle$ defined by the half-filled shell, they take on a simple form
 \[ {[\mathrm{GH}]}^{ L}_{m_\nu \, m_L} |0_N(i)\rangle = \left[ \boldsymbol{b}^\dagger(i)^{L+m_\nu} \boldsymbol{\tilde{b}}(i)^{L-m_\nu} \right]^{L}_{m_L}  |0_N(i)\rangle  \tag{c} \label{tGH}\]
 where the GH operators  $[\mathrm{GH}]^{ L}_{m_\nu \, m_L}$ carry the CF's angular momentum quantum numbers, $L$, $m_L$, and $m_\nu$,
 and act only in the electron space.  The vortex is unchanged across the multiplet.
 FQHE states of the same $N$ and opposite $m_\nu$, which correspond to the distinct fillings ${p \over 2p+1}$ and
 ${p \over 2p-1}$, are related by a simple exchange of electron creation and annihilation operators in their valence GH operators, 
 $\boldsymbol{b}^\dagger \leftrightarrow \boldsymbol{\hat{b}}$.   Thus we identify CF electron particle-hole conjugation as the
 associated symmetry operation.  States of filling $\nu={p \over 2p+1}$ and ${p \over 2p-1}$ with the same $N$
 have identical structures, and differ only by conjugation of
 their respective CFs.

 The algebraic correspondence apparent between the IQHE solutions $\mathcal{D}^{L}_{m_S \, m_L} (0,\theta, -\phi)$ and \eqref{tGH} is then explored.  The quantum numbers
 $(L,m_L)$ in both cases are associated with the angular momentum, and thus the location of either electrons or CFs on the sphere.  The second
 set $(L,m_\nu)$  appearing in \eqref{tGH} corresponds to a new SU(2) operator triad $(\hat{S}^\nu_0,\hat{S}^\nu_\pm)$,
 distinct from  
 $(\hat{S}_0, \hat{S}_\pm)$ of the IQHE:  the $\hat{S}^\nu_\pm$ generate changes in the filling, altering the distribution of a fixed
 number of quanta between the CF's electron and vortex spinors,
 while the $\hat{S}_\pm$ generate changes in the magnetic field.
 $\hat{L}^2=\hat{S}^{\nu \, 2}=L(L+1)$, showing the dual role of the eigenvalue $L$.  
 We also introduce operators to raise or lower the number of vortex quanta, $\boldsymbol{v}^\dagger$ and
 $\boldsymbol{\tilde{v}}$, and show that the angular momentum and $\nu$-spin operators can be written as bilinears in electron and vortex
 ladder operators.   Thus the FQHE is formulated as a two-component (electron and vortex) system governed by two independent
 and commuting SU(2) algebras, similar to the proton/neutron spin/isospin CF description of QCD.
 
 In Sec. V we discuss particle-hole (PH) symmetry, which relates FLL states of filling ${p \over 2p+1}$ and ${\bar{p} \over 2p+1}$, $\bar{p}\equiv p+1$, residing in
 the same magnetic field $m_S$, but with $N < \bar{N}$, constrained by $N+\bar{N}=2m_S+1$.    Thus to link their CFs 
 algebraically, we must identify new multiplets that preserve $m_S$, not $N$.  We show that the multiplet ladder operators 
 are  $\hat{S}^\nu_\pm$.  This leads us to a new set of $\nu$-spin multiplet
 operators GH$^\nu$ relating PH conjugate wave functions of distinct $N$,
 \begin{eqnarray*}
 \Psi^{N \, L}_{m_\nu \, m_L} &=& [ \mathrm{GH^\nu}]^{ L}_{m_\nu \, m_L} |0_{N_{1 \over 2}}(i) \rangle  \\
 &=& \left[ \boldsymbol{d}^\dagger(i)^{L} \otimes \boldsymbol{\tilde{d}}(i)^{L} \right]^{L}_{m_\nu \, m_L} |0_{N_{1 \over 2}}(i) \rangle ~~~~~~~~~~~~~\mathrm{(d)}
 \end{eqnarray*}
 written with respect to a vacuum defined by the half-filled shell.
 The operator $\boldsymbol{d}^\dagger$ is a two-component ladder operator analogous to $\boldsymbol{c}^\dagger$ of
  \eqref{tIQHE}, but operating in $\nu$-spin space rather than magnetic field space, that is, $\hat{S} \rightarrow \hat{S}^\nu$.
  
  The generated CFs, for $\nu<{1 \over 2}$,  are electron-particle, vortex-hole excitations of the half filled shell, with the
  particle-hole number difference 0 (1) for integer (half-integer) angular momentum.   Thus we find that PH symmetry
  is manifested in CF representations as an electron-vortex exchange symmetry:  such an exchange converts $N$ CFs
  of  state $\Psi$ occupying subshells $\mathcal{I}=1, \dots, p$ into the $N$ CFs of conjugate $\nu>{1 \over 2}$ state,
  $\bar{\mathcal{I}}=2,\dots,\bar{p}=p+1$.
  However there are two interesting twists.  This symmetry is a mirror symmetry, despite
  the fact that $m_\nu=1+|\bar{m}_\nu|$, because the PH conjugate state related to $\Psi$ is shifted,
  \[ \Psi \leftrightarrow \hat{S}^\nu_- \bar{\Psi}   \]
  Second, the algebra places no constraint on the $\bar{N}-N$ CFs of the $\bar{\Psi}$ state occupying the lowest subshell,
  $\bar{\mathcal{I}}=1$, because that subshell is annihilated by $\hat{S}^\nu_-$. 
  
  This pattern is familiar from supersymmetric quantum mechanics and leads us to a discussion of the CF effective
  Hamiltonian in Sec. VI.  We argue that the Pauli Hamiltonian is
 \[ H_0^\mathrm{eff} = \hbar \omega_{Coul} \left[\epsilon_1 + {3 \over 2} \left(\epsilon_{{1 \over 3}} - \epsilon_1 \right) {1 \over S} \hat{S}^\nu_+ \hat{S}^\nu_- \right]  \]
 where $\epsilon_1=-\sqrt{{\pi \over 8}}$ and $\epsilon_{1 \over 3}$ are the average single-electron energies at $\nu=1$ and ${1 \over 3}$,
 respectively, and $\hbar \omega_{Coul}$ is the Coulomb energy scale ${\alpha \hbar c \over a_0}$, with $a_0$ the cyclotron
 radius.  This Hamiltonian is isospectral, with a form familiar from supersymmetric quantum mechanics, and also very similar to
 the Pauli Hamiltonian \eqref{tIQHEH} of the IQHE.  We demonstrate its consistency with the constraints PH symmetry places on the energies
 of PH conjugate states.   This Hamiltonian identifies the $\bar{\mathcal{I}}=1$ subshell of the ${\nu}>{1 \over 2}$ state as the zero mode,
 which $\hat{S}^\nu_-$ annihilates.  The zero mode mass is $\hbar \omega_{Coul} \epsilon_1$.
 
The  Pauli equation can be linearized, producing a
 Dirac Hamiltonian for the CFs of the FQHE that links the four components of two the mirror symmetric pairs
  \[ \Psi \leftrightarrow \hat{S}^\nu_- \bar{\Psi}~~~~~~\hat{S}^\nu_- \Psi \leftrightarrow \bar{\Psi}^P  \]
  where $\bar{\Psi}^P$ is the $N$-CF wave function obtain by restricting $\bar{\Psi}$ to its upper $p$ subshells.
 
 In Sec. VII we summarize our results and describe
 additional directions that might be explored.
 
\section{Spherical Geometry}
We first describe the noninteracting problem in spherical geometry \cite{Haldane} -- electrons confined
to the surface of a sphere, with the magnetic field generated by a Dirac monopole at the center.  The discussion establishes the algebraic 
framework for subsequent discussions of CFs.  Our treatment is most similar to that of Greiter \cite{Greiter},
extended to emphasize operator $SU(2) \times SU(2)$ labels and connections to underlying creation operators.

The important advantage of spherical geometry arises in the interacting problem: translationally invariant states of
uniform one-body density correspond uniquely to many-electron states
of total angular momentum $L$=0.  It is thus not surprising that CFs turn out to be fermion spinors carrying 
multiple angular momentum quantum labels,
from which scalar many-body states can be constructed by filling shells.  Note that spherical solutions
can be immediately written in planar geometry, following \cite{GH2}, where simple rules for mapping spherical
solutions to planar ones are provided, based on the correspondence $(L_x,L_y) \leftrightarrow (p_x,p_y)$
(connected to the Cayley-Klein parameters discussed below).
This procedure produces planar wave functions that are translationally invariant, while still allowing use of the
simple but overcomplete coordinates $z_i=x_i+i y_i$.

The monopole quantization condition requires the total magnetic flux to be $\Phi=2m_S \Phi_0$, where $2m_S$ is
the number of monopole quanta and thus an integer, and $\Phi_0=h c/q$ the elementary unit of flux. The noninteracting single-electron Hamiltonian is
\begin{align}
\label{eq:Ham0sp}
\hat{H}_0 = {1 \over 2m_e\, m_S  \, a_0^2}  \left| \vec{r} \times \left({\hbar \over i} \vec{\nabla} - {q \over c} \vec{A}  \right)\right|^2 
= {\omega \over 2m_S  \hbar} \vec{\Lambda}^2 
\end{align}
where $m_e$ is the electron mass,  $\vec{\Lambda}=\vec{r} \times \left({\hbar \over i} \vec{\nabla} - {q \over c} \vec{A} \right)$ is the dynamical
angular momentum, $\omega=q B/m_ec = \hbar/m_e a_0^2$ is the cyclotron frequency, and
$\vec{\nabla} \times \vec{A} = B \hat{\Omega}$ where $\hat{\Omega} \equiv \vec{r}/R$.  The 
guiding-center angular momentum operators $\vec{L}=\vec{\Lambda}+ \hbar S \hat{\Omega}$
satisfy the commutation relations $[L_i,L_j]=i \epsilon_{ijk} L_k$.  As $\vec{\Lambda}$ is normal to the
surface while $\hat{\Omega}$ is radial, $\hat{\Omega} \cdot \hat{\Lambda} =\vec{\Lambda} \cdot \hat{\Omega}=0$ 
and $\vec{L} \cdot \hat{\Omega} = \hat{\Omega} \cdot \vec{L} = \hbar m_S$.  These relations 
give $\vec{\Lambda}^2 = \vec{L}^2 - \hbar^2 m_S^2=\hbar^2( L(L+1)-m_S^2)$.
The Landau level eigenvalues and normalized wave functions are
\begin{eqnarray}
&& E = {\hbar \omega \over 2m_S} (L(L+1)-m_S^2) \nonumber \\
&&\Psi^L_{m_S \, m_L} = \sqrt{{2L+1 \over 4 \pi}}\mathcal{D}^L_{m_S \, m_L}(0,\theta,-\phi) \nonumber \\[1ex]
 &&-L \le m_S, \,m_L \le L
\end{eqnarray}
where $\mathcal{D}$ is the Wigner D-function (rotation matrix).  The solutions carry quantum numbers in
two SU(2) spaces, $(L,m_L)$ and $(S,m_S)$, but with $L=S$.

We employ the rotation matrices of \cite{Varshalovich},
which have the property that they transform as good spherical tensors in both magnetic indices,
and thus satisfy simple recursion formulas involving Clebsch Gordan coefficients.
The elementary spinor basis for these solutions are the unitary $2 \times 2$ matrices connected 
with the Cayley-Klein parameters \cite{Varshalovich}.  They define the basis for the 
spherical $L=S=\textstyle{1 \over 2}$ creation operators
\begin{widetext}
\begin{equation}
\boldsymbol{c}^\dagger_{m_S  m_L} \equiv \left( \begin{array}{c} \boldsymbol{c}^\dagger_{~{1 \over 2}~{1 \over 2}} \\ \boldsymbol{c}^\dagger_{~{1 \over 2}-{1 \over 2}} \\ \boldsymbol{c}^\dagger_{-{1 \over 2}~{1 \over 2}} \\ \boldsymbol{c}^\dagger_{-{1 \over 2}-{1 \over 2}} \end{array} \right)
\equiv  \left( \begin{array}{c} \boldsymbol{b}^\dagger_{~{1 \over 2}} \\ \boldsymbol{b}^\dagger_{-{1 \over 2}} \\ \boldsymbol{a}^\dagger_{~{1 \over 2}} \\ \boldsymbol{a}^\dagger_{-{1 \over 2}} \end{array} \right)
 \rightarrow \left( \begin{array}{c} \mathcal{D}^{~{1 \over 2}}_{~{1 \over 2}~{1 \over 2}}(0,\theta,-\phi) \\[1.7ex] \mathcal{D}^{~{1\over 2}}_{~{1 \over 2}-{1 \over 2}}(0,\theta,-\phi) \\[1.7ex] \mathcal{D}^{~{1 \over 2}}_{-{1 \over 2}~{1 \over 2}}(0,\theta,-\phi) \\[1.7ex] \mathcal{D}^{~{1 \over 2}}_{-{1 \over 2}-{1 \over 2}}(0,\theta,-\phi) \end{array} \right)  
 =  \left( \begin{array}{c} \cos{{\theta \over 2}} e^{i \phi/2} \\[1.7ex]-\sin{{\theta \over 2}} e^{-i \phi/2}  \\[2ex] \sin{{\theta \over 2}} e^{i \phi/2}  \\[2ex] \cos{{\theta \over 2}} e^{-i \phi/2} \end{array} \right) 
 \equiv  \left( \begin{array}{c}~ u_{~{1 \over 2}} \\[2ex] ~u_{-{1 \over 2}}  \\[2ex] -u^*_{-{1 \over 2}}  \\[2ex] ~u^*_{~{1 \over 2}} \end{array} \right) 
 =   \left( \begin{array}{c}~ ~u \\[2ex] ~\textstyle{-}v  \\[2ex] ~{v}^*  \\[2ex] ~{u}^* \end{array} \right) 
\end{equation}
\end{widetext}
The operators (left) and the corresponding elementary spinors (right) are given, along with their relationship to Haldane's $u$ and $v$.
The annihilation operators are
\begin{equation}
\boldsymbol{\tilde{c}}_{m_S \, m_L} \equiv \left( \begin{array}{c} \boldsymbol{\tilde{c}}_{~{1 \over 2}~{1 \over 2}} \\ \boldsymbol{\tilde{c}}_{~{1 \over 2}-{1 \over 2}} \\ \boldsymbol{\tilde{c}}_{-{1 \over 2}~{1 \over 2}} \\ \boldsymbol{\tilde{c}}_{-{1 \over 2}-{1 \over 2}} \end{array} \right)
\equiv  \left( \begin{array}{c} \boldsymbol{\tilde{a}}_{~{1 \over 2}} \\ \boldsymbol{\tilde{a}}_{-{1 \over 2}} \\ \boldsymbol{\tilde{b}}_{~{1 \over 2}} \\ \boldsymbol{\tilde{b}}_{-{1 \over 2}} \end{array} \right)
 \rightarrow \left( \begin{array}{c} {d/du^*_{1 \over 2}} \\[2ex]{d/d u^*_{-{1 \over 2}}} \\[2ex] -{d/du_{-{1 \over 2}}} \\[2ex] {d/ du_{{1 \over 2}}} \end{array} \right)\\
\end{equation}
The Cayley-Klein parameters
define stereographic projection onto the plane, and thus are connected with transformations that relate
spherical results to the planar spinor products appearing in the wave functions of \cite{GH2}, as mentioned previously. 
The nonzero commutation relationships among the ladder operators are
\begin{flalign}
&[\boldsymbol{\tilde{b}}_{m_L}, \boldsymbol{b}^\dagger_{m_L^\prime} ] = \delta_{m_L,-m_L^\prime} (-1)^{{1 \over 2} +m_L} \nonumber \\
&[\boldsymbol{\tilde{a}}_{m_L}, \boldsymbol{a}^\dagger_{m_L^\prime} ] = \delta_{m_L,-m_L^\prime} (-1)^{{1 \over 2} -m_L} \nonumber \\
&[\boldsymbol{\tilde{c}}_{m_S \, m_L}, \boldsymbol{c}^\dagger_{m_S^\prime \, m_L^\prime} ] = \delta_{m_S \, -m_S^\prime} \delta_{m_L \, -m_L^\prime} (-1)^{{1 \over 2} - m_S+{1 \over 2} -m_L}
\end{flalign}
The four-component operators $\boldsymbol{c}^\dagger$, $\boldsymbol{\tilde{c}}$ reside in a direct product angular-momentum/magnetic-field 
space associated with the operators $\hat{L}$ and $\hat{S}$ defined below, while $\boldsymbol{b}^\dagger$ and $\boldsymbol{\tilde{b}}$ as
well as  $\boldsymbol{a}^\dagger$ and $\boldsymbol{\tilde{a}}$ operate in angular momentum space.

One can form various bilinear operators with simple transformation properties by combining the 
ladder operators into spherical tensors with definite $S,L$.  An operator that transforms as an angular momentum
vector and a magnetic field scalar is
\begin{eqnarray}
\hat{L}_{1m_L} &=& \hbar [\boldsymbol{c}^\dagger \otimes \boldsymbol{\tilde{c}} ]_{S=0 m_S=0;L=1 m_L} \nonumber \\
&=& {\hbar \over \sqrt{2}} \left( [\boldsymbol{b}^\dagger \otimes \boldsymbol{\tilde{b}}]_{1 m_L} -  [\boldsymbol{a}^\dagger \otimes \boldsymbol{\tilde{a}}]_{1 m_L} \right) \nonumber \\
&\equiv& \hat{L}_{b \, 1m_L} + \hat{L}_{a \, 1m_L}
\label{eq:angmom}
\end{eqnarray}
which we recognize as the angular momentum operator.  Here $\otimes$ denotes the standard tensor product,
taken either in the combined angular momentum/magnetic field space (first line), or (second line)
just in angular momentum, as indicated by the quantum labels on the product.  Writing out the component form for $\hat{L}_{1 \, 0}=\hat{L}_z$, 
\begin{flalign}
 \hat{L}_z = {\hbar \over 2} \left( \boldsymbol{b}^\dagger_{1 \over 2} \boldsymbol{\tilde{b}}_{-{1 \over 2}} +\boldsymbol{b}^\dagger_{-{1 \over 2}} \boldsymbol{\tilde{b}}_{{1 \over 2}}- \boldsymbol{a}^\dagger_{1 \over 2} \boldsymbol{\tilde{a}}_{-{1 \over 2}} -\boldsymbol{a}^\dagger_{-{1 \over 2}} \boldsymbol{\tilde{a}}_{{1 \over 2}} \right)~~
\end{flalign}
one can then convert to Haldane's notation
\[ \hat{L}_z ={\hbar \over 2} \left( u {d \over du} - v {d \over dv}+{v}^* {d \over d {v}^*} -u^* {d \over d {u}^*} \right) \]
though the manifest $(S,m_S; \, L,m_L)=(0,0;1,0)$ character of the operator is then lost.
Using the standard raising and lowering operators
\begin{eqnarray}
\hat{L}_+ &\equiv& -\sqrt{2}\hat{L}_{11}= \hat{L}_x +i \hat{L}_y \nonumber \\
\hat{L}_- &\equiv& \sqrt{2}\hat{L}_{1-1}= \hat{L}_x -i \hat{L}_y 
\end{eqnarray}
one can easily verify that $\boldsymbol{b}$, $\boldsymbol{a}$, and $\mathcal{D}^L_{m_S \, m_L}$ all transform under $\hat{L}$ as standard
angular momentum spinors, e.g.,
\begin{flalign}
&~~~~\hat{L}_z \,  \mathcal{D}^L_{m_S \, m_L}(0,\theta,-\phi) = \hbar m_L  \mathcal{D}^L_{m_S \, m_L}~~~~~~~~~~~~~~~~~~~~~~~~~~~~~~~~~~~~ \nonumber \\
&~~~~\hat{L}_+ \mathcal{D}^L_{m_S \, m_L}(0,\theta,-\phi) = \nonumber \\
&~~~~~~~~~~~~~~\hbar  \sqrt{(L-m_L)(L+m_L+1)} \mathcal{D}^L_{m_S \, m_L+1} \nonumber \\
&~~~~\hat{L}_- \mathcal{D}^L_{m_S \, m_L}(0,\theta,-\phi) = \nonumber \\
&~~~~~~~~~~~~~~\hbar  \sqrt{(L+m_L)(L-m_L+1)} \mathcal{D}^L_{m_S \, m_L-1}
\end{flalign}
while
\begin{equation}
[\hat{L}_i,\hat{L}_j]= i \hbar \, \epsilon_{ijk}  \hat{L}_k
\end{equation}
for the Cartesian components.

The analogous operator that transforms as a magnetic field space vector but an angular momentum scalar is
\begin{eqnarray}
\hat{S}_{1m_S} &=& [\boldsymbol{c}^\dagger \otimes \boldsymbol{\tilde{c}} ] _{S=1 m_S;L=0 m_L=0}.
\end{eqnarray}
One finds
\begin{eqnarray}
\hat{S}_{1 \, 0} \, &\equiv& \hat{S}_z = {1 \over 2} \left( \boldsymbol{b}^\dagger \odot \boldsymbol{\tilde{b}}+ \boldsymbol{a}^\dagger \odot \boldsymbol{\tilde{a}} \right) \nonumber \\
\hat{S}_+ &\equiv& -\sqrt{2}\hat{S}_{11}= \hat{S}_x +i \hat{S}_y = - \boldsymbol{b}^\dagger \odot  \boldsymbol{\tilde{a}}  \nonumber \\
\hat{S}_- &\equiv& \sqrt{2}\hat{S}_{1-1}= \hat{S}_x -i \hat{S}_y =  \boldsymbol{a}^\dagger \odot  \boldsymbol{\tilde{b}}
\end{eqnarray}
Here we have defined the scalar product of two spherical tensors of rank $J$ as
\[ A_J \odot B_J = \sum_M (-1)^{J-M} A_{JM} B_{J-M} \]
(which for $J=1$ differs by a sign from the standard vector dot product).
One can then verify that $\boldsymbol{b}$, $\boldsymbol{a}$, and $\mathcal{D}^L_{m_S \, m_L}$ all transform conventionally under $\hat{S}$, e.g.,
\begin{flalign}
&\hat{S}_0 \, \mathcal{D}^L_{m_S \, m_L}(0,\theta,-\phi) = m_S \mathcal{D}^L_{m_S \, m_L}  \nonumber \\
&\hat{S}_+ \mathcal{D}^L_{m_S \, m_L}(0,\theta,-\phi) = \nonumber \\
&~~~~~~~~~~~~~~ \sqrt{(L-m_S)(L+m_S+1)} \mathcal{D}^L_{m_S+1 \, m_L} \nonumber \\
&\hat{S}_- \mathcal{D}^L_{m_S \, m_L}(0,\theta,-\phi) = \nonumber \\
&~~~~~~~~~~~~~~ \sqrt{(L+m_S)(L-m_S+1)} \mathcal{D}^L_{m_S-1 \, m_L}
\end{flalign}
and that 
\begin{equation}
[\hat{S}_i,\hat{S}_j]= i \epsilon_{ijk}  \hat{S}_k~~~~~~~~~[\hat{L}_i,\hat{S}_j]=0
\end{equation}
for Cartesian components.
Haldane \cite{Haldane} considered $\hat{S}_0$, restricted to the FLL, and Greiter \cite{Greiter}
appears to have first utilized the
set $(\hat{S}_0,\hat{S}_+,\hat{S}_-)$ (though he may have misplaced a relative sign that then 
prevents grouping these operators into a tensor).

The Hamiltonian of Eq. (\ref{eq:Ham0sp}) can be rewritten in terms of $\hat{S}_\pm$, 
\begin{eqnarray}
\hat{H} &=& {\hbar \omega \over 2S}~ {1 \over 2} \left[ \hat{S}_+\hat{S}_- +\hat{S}_-\hat{S}_+ \right],
\label{eq:plusmi}
\end{eqnarray}
Greiter's main result \cite{Greiter}.  This form, a product of operators that raise and then lower (or the reverse)
the magnetic field by a unit, has a Dirac equation analog that was studied by Arciniaga and
Peterson \cite{Arciniaga}, and will be used later in this paper.  (See also \cite{Jellal}.)  We will later find the
the effective Hamiltonian for GH$^2$ CFs has a form very similar to Eq. (\ref{eq:plusmi}).

\section{GH$^2$ Composite Fermions}
The GH operators were originally derived by extending Laughlin's variational arguments.   Laughlin considered
the two-electron correlation function, identifying the translationally invariant ground states of maximum density that
exclude $p$-wave ($\nu=\textstyle{1 \over 3}$) or $p$- and $f$-wave ($\nu=\textstyle{1 \over 5}$) interactions.  While superficially one 
might interpret this construction in terms of short-range physics, its success derives from producing the quantum mechanical
analog of a scale-invariant wave function, consistent with the underlying Coulomb interaction.  GH recognized that the analogous scaling for
other fillings cannot be described in terms of just the two-electron correlation function: at densities 
beyond $\nu=\textstyle{1 \over 3}$, $p$-wave correlations must appear, and thus to avoid exacerbating the cost in energy of a local
overdensity, higher order correlations are necessary to keep other electrons from approaching that $p$-wave pair. The GH construction addressed this problem systematically, 
progressing through 3-, 4-, and higher-body correlations relevant to $\nu=\textstyle{2 \over 5}$, $\textstyle{3 \over 7}$, $\dots$.
The correlations among the relevant clusters were set up to allocate the available quanta in the most symmetric way, minimizing
the costs of the accompanying over- or under-densities.  Rotational invariance and homogeneity were then restored by 
antisymmetrizing over all such partitions.   The net result was an analytic FLL extension of Laughlin's wave function in which a second
quantum was added to Laughlin's $m$.  (Laughlin's wave function is build from pairwise antisymmetric correlations of the form $(u(i) \cdot u(j))^m$,
$m=1,3,5, ...$, yield states of $\nu={1 \over m}$.)  For fixed $m=3$ this second quantum number enumerates the series of wave
functions $\nu=\textstyle{1 \over 3}, \textstyle{2 \over 5}, \textstyle{4 \over 9}, \cdots, \textstyle{4 \over 7}, \textstyle{3 \over 5}, 
\textstyle{2 \over 3},1$.  

The GH operators were originally applied directly to the $N$-electron bosonic $\nu=\textstyle{1 \over 2}$ state, following the procedure
used by Jain, who employed multiply-filled IQHE states as an operator and numerial projection to generate
FLL results.  Recently it was noted that the GH operators could be applied in a simpler way, while retaining
consistency with the Laughlin wave function \cite{GH2}.  This second procedure exploits a
factorization of the $N$-electron bosonic $\nu=\textstyle{1 \over 2}$ state into a product of $N$ translationally invariant
scalars, formed from the tensor product of electron and vortex spinors.  
One can regard these scalars as the $\nu={1 \over 2}$ seeds 
for the CF intrinsic states.
The GH$^2$ CF electron-vortex structure is connected with certain symmetries the relate FQHE states of different fillings,
explored for the first time here.
While numerically the Jain and GH$^2$ constructions produce 
variational wave functions with comparably excellent overlaps with numerically-generated wave functions, the 
second procedure explicitly constructs the CFs and the noninteracting many-CF wave functions, which take the form of $p$ closed
CF subshells.  The technical details of the construction have already been described \cite{GH2}.
The current paper focuses on the quantum numbers of the CFs and the symmetries of the
associated many-electron wave functions.  The construction places the hierarchy of FQHE states into
multiplets, with states of different filling related through simple operator transformations.

\subsection{Forms of the GH$^2$ Wave Function}
The GH operators have
an $\ell s$ form, where $\ell$ and $s$ are associated with the number of ladder operators that are employed to create or destroy
electron quanta with respect to the half-filled vacuum.
As in other angular momentum settings, $\ell$ and $s$ can be integer or half integer, 
constrained by $(2\ell+1)(2s+1)=N$: the operators are closed $\ell s$ shells, and thus translationally invariant.   When applied
in the manner of GH$^2$, the resulting CF
wave functions take the form of
a single determinant,  filling $p=2s+1$ closed subshells, mapping the wave function into a noninteracting form. 
No similar form for the Jain
or Jain/GH wave functions has yet been derived.\\

 \noindent
 {\it The GH$^2$ $\ell s$ form:} The GH$^2$ wave function can be written in equivalent  $\ell s$ (native), $(\ell s)j$ (composite fermion),
and hierarchical forms, useful in understand the electron-electron correlation, single-particle-structure, and 
many-body correlation structure of the wave functions, respectively.  Laughlin's $\nu=1/3$ wave function,
\begin{eqnarray}
 \label{Laughx}
 &&\Phi_{\ell \, 0} = \sum_{M's} \epsilon_{M_1 \mycdots M_N}~ [u_1]^\ell_{M_1} \mycdots [u_N]^\ell_{M_N} \nonumber \\
 && ~~~\times  [u_1]^{N-1 \over 2} \odot [u_1 \mycdots u_N;\bar{u}_1]^{N-1 \over 2} \times  \mycdots \nonumber \\
 &&~~~\times  [u_N]^{N-1 \over 2} \odot [u_1 \mycdots u_{N};\bar{u}_{N}]^{N-1 \over 2} \nonumber
 \end{eqnarray}
 where $2 \ell$+1 = $N$, becomes the first member of the GH$^2$ $\ell s$ sequence,  (2$\ell$+1)(2$s$+1)=$N$,
 \begin{eqnarray} 
 \label{GH2}
 &&\Phi_{\ell \, s} = \sum_{m's \, q's} \epsilon_{M_1 \mycdots M_N} [u_1]^\ell_{m_1} \mycdots [u_N]^\ell_{m_N}  \nonumber \\
&& ~~~\times   \left[ [u_1]^{{N-1 \over 2}-s} \otimes  [u_1 \mycdots u_N;\bar{u}_1]^{N-1 \over 2} \right]^s_{q_1} \times \mycdots \nonumber \\
&&~~~\times \left[ [u_N]^{{N-1 \over 2}-s} \otimes [u_1 \mycdots u_{N};\bar{u}_N]^{N-1 \over 2} \right]^s_{q_N}  
 \end{eqnarray}
Here $[u_1 \mycdots u_{N};\bar{u}_i]^{N-1 \over 2}$ is the vortex spinor for $N$-1 electrons, the aligned coupling
of $N-1$ elementary spinors with the $i$th
electron omitted, thus forming a spinor of angular momentum $(N-1)/2$;
 $M_i\equiv \left\{m_i,q_i \right\}$; $-\ell \le m_i \le \ell$; $-s \le q_i \le s$; $\epsilon$ is the antisymmetric
tensor; $\odot$ and $\otimes$ are the spherical scalar and tensor products; and $2s$ and $2\ell$ are integers.
Hierarchy states ($\nu \le {1 \over 2}$) with $\nu$=${p /(2p+1)}$ $\equiv$
$(2s+1)/(4s+3)$=${1 \over 3},{2 \over 5}, {3 \over 7}, \mycdots$ correspond to fixed $s=0,{1 \over 2},1, \mycdots$,
respectively, with $\ell \ge s$ determining $N$.  Conjugate states ($\nu \ge{1 \over 2}$) with $\nu$=
$p/(2p-1)$ $\equiv$ ${(2\ell+1)/(4 \ell+1)}$=$ {1},{2 \over 3},{3 \over 5} \mycdots$ correspond to fixed $\ell= {0},{1 \over 2},{1,} \mycdots$, respectively, with $s \ge \ell$
determining $N$.  The closed-shell $\ell$ and $s$ structure guaratees that the many-body states have the required $L^T=0$.\\
 
 \noindent
 {\it The GH$^2$ $(\ell s)j$ CF form:}  Eq. (\ref{Laughx}) can be rewritten in its CF form by combining
 the two $u_i$ factors.  The Laughlin wave function becomes a single closed shell of 2$\ell$ +1 CFs
 \begin{eqnarray}
 \Psi^N_{ \ell \,  s=0 \, m}(1)= \left[ [u_1]^{{N-1 \over 2} + \ell} \otimes  [u_2 \mycdots u_N]^{N-1 \over 2} \right]^\ell_{m} 
  \label{Laugh2}
 \end{eqnarray}
 formed by taking the anti-aligned product of the electron and vortex spherical tensors.
 Eq. (\ref{GH2}) can be handled similarly, yielding
  \begin{equation}
 \Psi_{ \ell \, s \, m}^{N \,\mathcal{I}}(1)= \left[ [u_1]^{{N-1 \over 2} + \ell-s} \otimes  [u_2 \mycdots u_N]^{N-1 \over 2} \right]^{|\ell-s|+\mathcal{I}-1}_{m}  
  \label{GH2A}
 \end{equation}
 For $\nu=p/(2p+1) \le {1 \over 2}$ (so $\ell \ge s$), there are $p=2s+1$ filled subshells filled by their respective CFs,
  $\mathcal{I}=1, \mycdots, p$, forming the angular momentum tower  $L$=$\ell-s, \mycdots, \ell+s$.
 The number of subshells increases with $\nu$, reaching its maximum 
 at $\nu$=${1 \over 2}$ ($\ell$=$s$), where $p$=$\sqrt{N}$.  For $\bar{\nu}=p/(2p-1) \ge {1 \over 2}$
 ($s \ge \ell$) there are
 $p$=2$\ell$+1 subshells,  $\mathcal{I}=1, \mycdots, p$, forming the angular momentum tower  $L$=$s-\ell, \mycdots,s+\ell$.
The height of the tower decreases as $\bar{\nu} \rightarrow 1$, reducing to one shell at $\bar{\nu}$=1.
 
 The wave function for a given $\nu$ is formed from the product of the closed shells, antisymmetrized over exchange of CFs among the shells (that is, a single determinant is formed).
 The CF form of the GH$^2$ wave function is discussed in detail in \cite{GH2}.\\

  \noindent
 {\it The GH$^2$ hierarchical form:} The Laughlin wave function can be written as a product
 of $N$ translationally-invariant scalars, distributed over the sphere according to a pairwise spin-spin correlation
 \begin{eqnarray}
\Phi_{\ell \, 0} &=& \prod_{i=1}^N R(i)~ \prod_{i<j=2}^N ~{\Sigma}(i) \odot {\Sigma}(j) \nonumber \\
 R(i)&=& [u_i]^{N-1 \over 2} \odot [u_1 \cdots u_N;\bar{u}_i]^{N-1 \over 2} \nonumber \\
  \Sigma(i)_m &=& [u_i]^{1 \over 2}_m
 \end{eqnarray}
 The hierarchical GH$^2$ wave function emerges from $\ell s$  coupling before antisymmetrization, yielding the generalization (for $\nu<{1 \over 2}$)
  \begin{equation}
\Phi_{\ell \, s} = \prod_{I=1}^{2 \ell+1} R_s(I)~ \prod_{I<J=2}^{2 \ell+1} ~{\Sigma}_s(I) \odot {\Sigma}_s(J) 
\end{equation}
where the $N$ electrons have been partitioned into  $2\ell$+1=$N$/(2$s$+1) sets containing $2s$+1 equivalent electrons each, enumerated 
by $I,J,...$.  Denoting $I = \{i_1,\cdots,i_{2s+1} \}$,
\begin{eqnarray}
 &&R_s(I)=\sum_{m's} \epsilon_{m_1 \cdots m_{2s+1}} \times \nonumber \\
 &&~~~~~\left[ [u_{i_1}]^{{N-1 \over 2}-s} \otimes [u_1 \mycdots u_N;\bar{u}_{i_1}]^{N-1 \over 2} \right]^s_{m_1} \nonumber \\
 &&~~~~~ \cdots \left[ [u_{i_{2s+1}}]^{{N-1 \over 2}-s} \otimes [u_1 \mycdots u_N;\bar{u}_{i_{2s+1}}]^{N-1 \over 2} \right]^s_{m_{2s+1}}\nonumber \\
 && {\Sigma}_s(I)_m = \left[ u(i_1) \cdots u(i_{2s+1}) \right]^{2s+1 \over 2}_m
 \end{eqnarray}
 The scalar $R_s(I)$ is a closed shell of 2$s$+1 CFs.
 The wave function is obtained by antisymmetrizing over all partitions.
 
 In the hierarchical picture, the underlying objects distributed on the sphere are vortices of length $2L$=$2s+1$, where $L$ is the vortex angular momentum.  Attached to each vortex is a translational 
 invariant ($L$=0) intrinsic state involving $2s+1$ electrons.  The intrinsic state represents a slight overdensity.  These overdensities
 are then spaced over the plane by the vortex spin-vortex spin pairwise interaction, the same variational mechanism used by Laughlin to
 separate and space single electrons.  As the CF and hierarchical wave functions take on Laughlin's form at $\nu={1 \over 3}$,
 each can be viewed as a series in which Laughlin's state is the first member.
 
 The hierarchical nature of this wave function can be visualized by fixing $\ell$ and incrementing $s=0,{1 \over 2},\mycdots $,
 proceeding through the fillings $\nu={1 \over 3},{2 \over 5}, \mycdots$.  The form of the wave function remains fixed:
 $2\ell$+1 translationally invariant scalars distributed over the sphere by pairwise interactions.   
 It is readily seen that both the scalars $R_s$ and vortices $\Sigma_s$ satisfy simple recursion
 relations, with the vortex angular momentum increasing by $\Delta L$=${1 \over 2}$  in each step.  Thus, successively, the
 $2N$-electron $\nu$=${2 \over 5}$ wave can be built from the $N$-electron $\nu$=${1 \over 3}$ wave, the $3N$-electron
 $\nu={3 \over 7}$ from the $2N$-electron $\nu={2 \over 5}$, etc.  This matches the descriptions 
 of the hierarchical constructions proposed by Haldane and Halperin \cite{Haldane,Halperin}.  As there has been some
 debate over the equivalence of hierarchical and CF constructions \cite{deb1,deb2,deb3,deb4,deb5}, it is satisfying that the GH$^2$ wave
 functions can be written in equivalent CF and hierarchical forms.  The shell structure of the former and the clustering  of the latter
 are manifestations of a familiar choice -- couple or uncoupled representation -- one makes in treating
 the GH operator $\ell \, s$ algebra.  The subshell structure of the CF representation reflects
 the distinct energies one encounters when successively removing electrons
 from a correlated cluster.
 
 %The hierarchical form, prior to antisymmetrization, makes it transparent that a state of filling $p/(2p+1)$ consists
% of overdense sets of $p$ electrons, correlated with each other in the manner of Laughlin's wave function,
 %distributed as $1/(2p+1)$. 
 %The hierarchical form of the wave function is the more natural formulation to use if one is focused on
 %physics issues associated with the long-range multi-electron correlations induced by the 
% Coulomb interaction, while the CF representation is the more appropriate form if one is 
% interested in gaps and low-lying excitations. 
 %  Note that $\Sigma(i)$ is an eigenfunction of $L$:
% the angular momentum of the clusters is entirely associated with the cluster moving center
 %for wave functions within the FLL.   This is not necessarily the case for wave functions constructed in
 %higher LLs.   The GH$^2$ construction provides a very accurate
 %description of the $\nu$=$\textstyle{1 \over 2}$ fermionic state.  However it is 
 %known that the $\nu$=$\textstyle{5 \over 2}$ state has distinct properties, developing a gap.  We
 %will not explore this issue further here, but anticipate that the hierarchical 
 %representation will support richer angular momentum structure outside the FLL, one that allows
 %the internal wave function of each CF to deform, thereby allowing the system to further
 %minimize its energy by suitably coordinating such perturbations in neighboring sites \cite{GHH3}.
 
 \subsection{Physical interpretation of CFs}
 In Fig. 1.12 of Girvin's lecture notes \cite{GirvinLN} on the quantum Hall effect, there is a visual representation of the
 correlated $m=3$ (so $\nu={1 \over 3}$) Laughlin state, where a snapshot of electron positions taken during a Monte Carlo simulation of a
 1000-particle trial wave function
 is contrasted with the corresponding Poisson distribution of 1000 uncorrelated electrons.  This graphic was itself
 inspired by an earlier and similar figure of Laughlin \cite{Laugh2}.  It is immediately apparent that, in the correlated case,
 the electrons are more uniformly distributed in the plane, and in particular, that the two-particle correlation function
 is highly constrained, keeping electrons separated.
 
 The CFs of Eq. (\ref{GH2A}) are the algebraic manifestation of this correlation.  It the case of the $m=3$ Laughlin state,
 it states that arbitrary positioning of $N$ electrons among $2S+1$ magnetic states on the sphere, $S=\textstyle{3(N-1) \over 2}$, 
 has instead been replaced by a closed-shell distribution of $N$ CFs
 of reduced angular momentum $\ell=\textstyle{N-1 \over 2}$.  Thus the ``centers" are uniformly distributed over the sphere according
 to this reduced angular momentum.
 The reduced angular momentum indicates the CF is a extended object, so that fewer of them cover the surface,
 measured relative to bare electrons.
 
 The CF corresponding to electron 1 carries with it an intrinsic wave function -- an angular momentum scalar -- that in 
 the case of the Laughlin $m=3$
 state, consists of the product $\Pi_{i \ne 1} u(1) \cdot u(i)$.  As a scalar, the intrinsic
 wave function is translationally invariant and thus independent of the electron's location.  It represents an unusual kind
 of correlation function, not focused on short-range physics, but instead building into each CF 
 a correlation that increases the spacing between CF 1 and all other CFs,
 regardless of whether they are nearby or on the opposite side of the sphere.   At short length scales, however, this correlation
 does encase the negative electron in a region of surrounding positive charge:
 the absence of neighboring electrons produces a region where the neutralizing positive background dominates.
 
 For the more complicated states of the hierarchy, new CFs arise, while the physics gently adjusts.   In the case of the $\nu=\textstyle{2 \over 5}$
 state, in addition to the $\mathcal{I}=1$ Laughlin CF subshell, there is a $\mathcal{I}=2$ subshell where $L$ has been incremented by one unit.
 Thus, with respect to the Laughlin subshell, the second subshell can accommodate two additional CFs, implying a slight increase
 in the CF density.  This indicates that this CF is slightly more compact than Laughlin's.  
  As we will see in detail later, this angular momentum increase comes from breaking one of the intrinsic-state
 antisymmetric pairs, replacing it with a symmetric pair which then must align with the electron spinor, generating the angular momentum increment. 
  Electron and vortex 
 angular momentum are no longer completely opposed.  The absence of an antisymmetric pair allows some neighboring electron
 to more closely approach electron 1.  That is, there is a nonzero probability for two electrons to be in a relative p state.  On average the positively charged envelope around the center has been
 slightly reduced in size:  the various effects (the addition of a new CF subshell (incrementing $p$), a more compact CF, 
 an incremented CF $L$,  and the resulting higher occupancy of that closed shell) naturally correlate.
 This process repeats for each increment in $p$.   Each step in the CF angular momentum tower represents the point where 
 a higher density has been reached in which a unique translationally invariant state can be created as $p$ closed shells,
 with successive shells characterized by fewer preserved scalar pairs, a necessary consequence of the increased density.
 
 The standard picture of a CF as an electron coupled to two units of magnetic flux confuses this simple picture:  when two Laughlin CFs
 1 and 2 interact, of course, each carries a scalar intrinsic state, and thus in combination the intrinsic states
 contribute two powers of $u(1) \cdot u(2)$ to the corresponding electron-electron correlation function.   But each of the CFs 
 (for $m=3$ hierarchy states) contributes only a single power.
 The CFs remain fermions under this assignment, with the same exchange properties as the electrons from which they are
 derived.  That antisymmetry generates an additional power of $u(1) \cdot u(2)$ in the correlation.  The GH$^2$ CFs
 algebraically encapsulate the physics Girvin captured visually in his figure.
 
 One might be concerned that the allocation of quanta separately to electron and vortex spinors is arbitrary.
 It is helpful to keep in mind the familiar example of the nucleon as the composite fermion of QCD:  strongly interacting
 systems of interacting colored quarks and gluons are fully equivalent at low energy to an effective theory described in terms of
 colorless protons ($uud$) and neutrons ($udd$), CFs that capture in their masses 99\% of the interaction energy \cite{Weinberg}.  
The partitioning of electron quanta into separate electron and vortex spinors to form CFs
 is analogous to the apportioning of $u$ and $d$ quarks among protons and neutrons, though in the FQHE states, creating CFs of minimum
 angular momentum guide the organization, rather than color.   In QCD 
 the nucleon CFs carry a second magnetic quantum number, magnetic isospin $m_\tau$, describing the allocation of $u$ and $d$ quarks in the
 nucleon: four states of the $L$=${1 \over 2}$ nucleon are distinguished by the SU(2) magnetic labels $m_L$ and $m_\tau$.     
 There is a corresponding second magnetic quantum number, $\nu$-spin $m_\nu$,
 that we introduce below for the FQHE, describing the allocation of quanta between electron and vortex spinors.  Its consequences are
 quite similar
 to those of isospin, allowing us the organize many-electron states into $\nu$-spin multiplets, and making the 
 algebraic relationships among states of different filling very apparent.
 
  \section{Mirror $\nu$-spin symmetry and FQHE multiplet structure}
  Here we introduce $\nu$-spin and its associated  magnetic quantum number $m_\nu = \ell-s$, 
  useful in bringing out symmetries around the half-filled
  shell (where $m_\nu=0$).  The CFs become
     \begin{eqnarray}
   && \Psi^{N  L \, \mathcal{I}}_{ m_\nu \, m_L}(1) \nonumber \\
   && ~~=\left[ [u_1]^{{N-1 \over 2} +m_\nu} \otimes [u_2 \cdots u_N]^{N-1 \over 2}\right]^{L=|m_\nu| +\mathcal{I}-1}_{m_L} \nonumber \\
    &&~~= \left[ [u_1]^{{m_S+m_\nu \over 2}} \otimes [u_2 \cdots u_N]^{m_S-m_\nu \over 2}\right]^{L=|m_\nu| +\mathcal{I}-1}_{m_L} 
  \label{simple}
\end{eqnarray}
   using $N-1=m_S-m_\nu$.  ($\mathcal{I}$ is a redundant label, clearly, but we occasionally will include it
   to make discussions clearer.)
   One notices a similarity between the FLL CF
   and multi-LL IQHE single-particle solutions
   \[  \Psi^{N  L \, \mathcal{I}}_{m_\nu \, m_L} \leftrightarrow \mathcal{D}^L_{m_S \, m_L} \]
   As in the noninteracting case, it appears CFs carry two magnetic quantum numbers,
   associated with the $z$-components of angular momentum and $\nu$-spin:  
   we found above that $L \ge |\ell-s|=|m_\nu|$.   
   $m_\nu$ measures the difference between the number of quanta carried by the CF's electron
   and vortex,  $(n_e-n_v)/2$.  Thus $m_\nu=0$ labels the $\nu=\textstyle{1 \over 2}$
   case where the electron and vortex spinors are of equal length, and CFs transform into
   themselves under electron-vortex exchange.
  
  Below we arrange the states of the FQHE according to $m_\nu$, to show the simple multiplet
  pattern that emerges, mirror symmetric around the half-filled shell.  We will do so under the constraint
  that $N$ is constant, yielding one symmetry relating states of different filling; in a later section, we repeat this 
  process keeping the magnetic field $m_S$ fixed, yielding new multiplets and a second
  important symmetry relation.  In both cases the operator triad $(\hat{S}_0^\nu,\hat{S}_\pm^\nu)$, scalars with respect to angular momentum,
  play a crucial role.   The similarities between the noninteracting multi-shell IQHE and CF
  descriptions of the interacting FLL FQHE arise from the algebraic similarities between $(\hat{S}_0,\hat{S}_\pm)$ and
  $(\hat{S}^\nu_0,\hat{S}^\nu_\pm)$.   In our view, the confusion seen in the literature over issues like Jain's projection, Coulomb vs. magnetic energy
  scales, etc., derives from conflating these two operator sets, which describe entirely different physics:
  $\hat{S}$ governs the angular momentum around the cyclotron centers that accompanies excitation
  into higher LLs, for noninteracting electrons, while $\hat{S}^\nu$ governs the
  allocation of quanta between electrons and vortices in CF representations of FLL FQHE states. 
  
\begin{figure*}[ht]   
\centering
\includegraphics[width=1.0\textwidth]{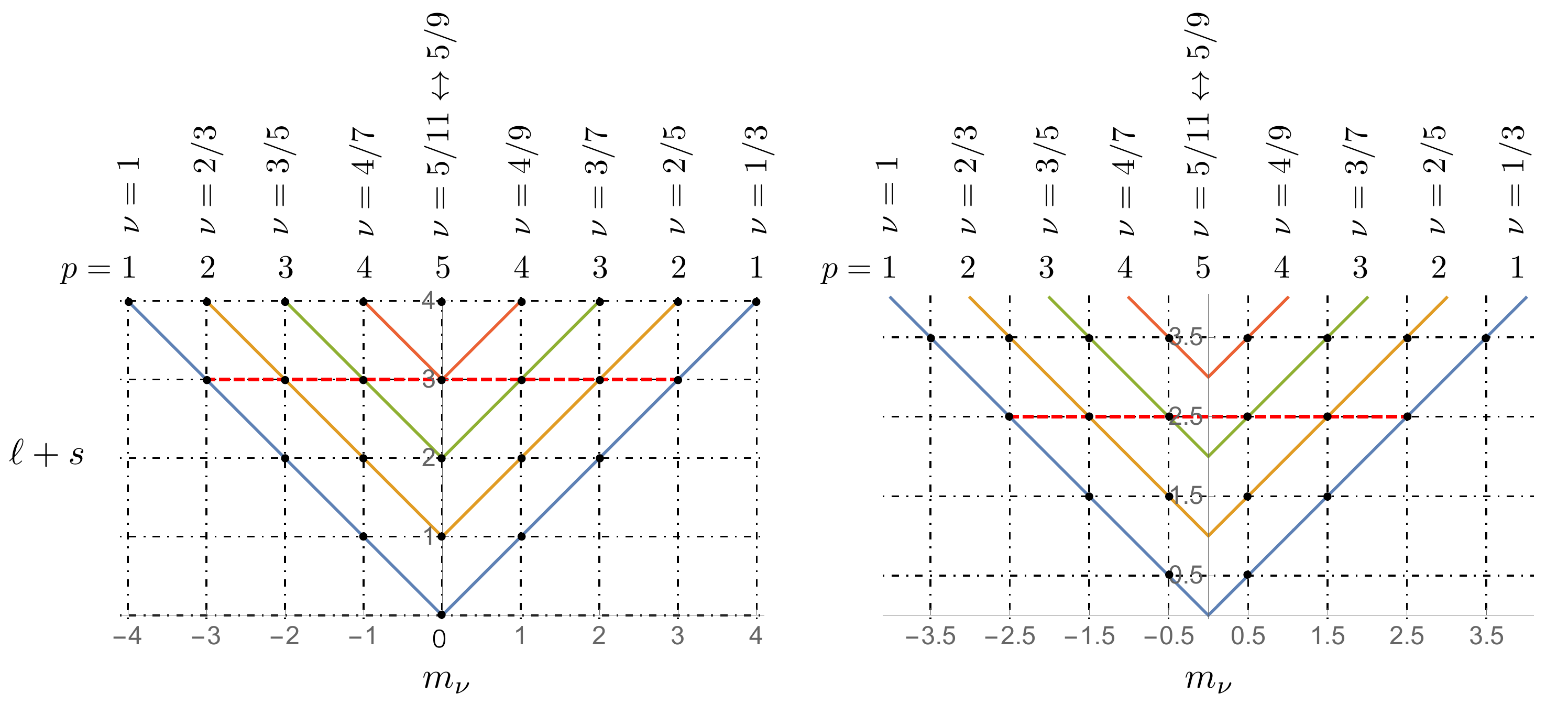}
\caption{The Hilbert space of GH$^2$ FQHE CFs
($m=3$ hierarchy).  There are two cases corresponding 
to integer (left) and half-integer (right) $L,m_L,m_\nu$.  The rotation from the $(\ell,s)$ GH variables to $\ell+s$ and $m_\nu=\ell-s$ 
makes the mirror symmetry in $m_\nu$ apparent: $N$ and $p$ are
unchanged when $m_\nu \leftrightarrow -m_\nu$. 
Trajectories of constant $p$ are indicated, corresponding to incrementing $N$ in states of a given filling.
For a specified $(N,p)$ the two values $\pm |m_\nu|$ correspond to mirror states of distinct filling,
$\nu=p/(2p+1)$ ($m_\nu>0$) and $p/(2p-1)$ ($m_\nu<0$).  
The red dashed
lines are the trajectories of constant $\ell+s$ corresponding to panels d) and c) in  Figs. \ref{Fig:Even} and  \ref{Fig:Odd}, respectively.}
\label{Fig:sl}
\end{figure*}  

\subsection{FQHE $\nu$-spin multiplets}

\subsubsection{CF quantum numbers}
The Hilbert space of $m=3$ FQHE states -- the many-electron
states formed from filling $p$ subshells with their respective CFs -- is a two-dimensional discrete grid, corresonding
to the GH variables $(\ell,s)$, which
take on all integer and half-integer values, with $\ell \ge0$ and $s \ge 0$.  To make use of $m_\nu$, and noting $|\ell-s| \le L \le \ell+s$,
one should rotate the $(\ell,s)$ axes by 45$^\circ$ to $\ell+s$ and $\ell-s = m_\nu$.  As in other
angular momentum problems, this space separates into two, corresponding to
$\ell+s$ and $\ell-s$ being either both integer, or both half integer.  This is depicted in
Fig. \ref{Fig:sl}.  The electron number $N$ and $p$ are then
easily found to be
\begin{eqnarray}
N&=& (\ell+s+1)^2 - m_\nu^2=p(p+2|m_\nu|) \nonumber \\
p&=& (\ell+s+1) - |m_\nu| = \sqrt{N+m_\nu^2}-|m_\nu|
\label{eq:relations}
\end{eqnarray}
As $N$ and $|m_\nu|$ determine $p$, and as the specific value of  $\mathcal{I} \in \{1, \cdots, p\}$ 
is determined by $L$, we find that $N,L,m_L,m_\nu$ are a complete set of CF eigenvalues.

The mirror operation corresponds to 
\[ (\ell+s, m_\nu) \leftrightarrow (\ell+s,-m_\nu). \]
As $N$ and $p$ do not depend on the sign of $m_\nu$, we see that mirror symmetry relates
pairs of states of the same electron number $N$ and the same $p$.  But the filling $\nu$ is
double valued in $p$.  For large $N$ and fixed $p$ \cite{GH2},
\[ \nu = {p \over 2p+\mathrm{sign}[m_\nu]} ,\]
Thus mirror symmetry in $m_\nu$ connects the FQHE states
\begin{equation}
 \Phi \left(N,p,\nu={p \over 2p+1} \right) \leftrightarrow \bar{ \Phi}\left(N,p,\bar{\nu}={p \over 2p-1}  \right) ,
\end{equation}
 with the left-hand (right-hand) side corresponding to $m_\nu$ positive (negative).
  
  Mirror symmetry in $m_\nu$ links the conjugate pairs ${1 \over 3} \leftrightarrow 1$,
  ${2 \over 5} \leftrightarrow {2 \over 3}$, ${3 \over 7} \leftrightarrow {3 \over 5}$, $\dots$,
  of the same electron number.  It states that such states have identical many-body 
  structures: the same number of subshells $p$, $\mathcal{I}=1,\cdots, p$,
   filled by their respective CFs.   The states differ only in their CFs,
   which transform as $\overline{\mathrm{CF}}(m_\nu) = \mathrm{CF}(-m_\nu)$.  That is
   their CFs are members of the same magnetic multiplet, carrying opposite $m_\nu$.

  The ``pivot" of the mirror symmetry is $m_\nu=0$, where $N=p^2$, $p=1,2,\cdots$, and where many-electron
  states and their CFs carry
  nominally different filling labels $\nu=p/(2p+1)$ and $\bar{\nu}=p/(2p-1)$, but are in fact identical.
  That is, these states and their individual CFs are self-conjugate.
  
  \begin{figure*}[ht]   
\centering
\includegraphics[width=1.0\textwidth]{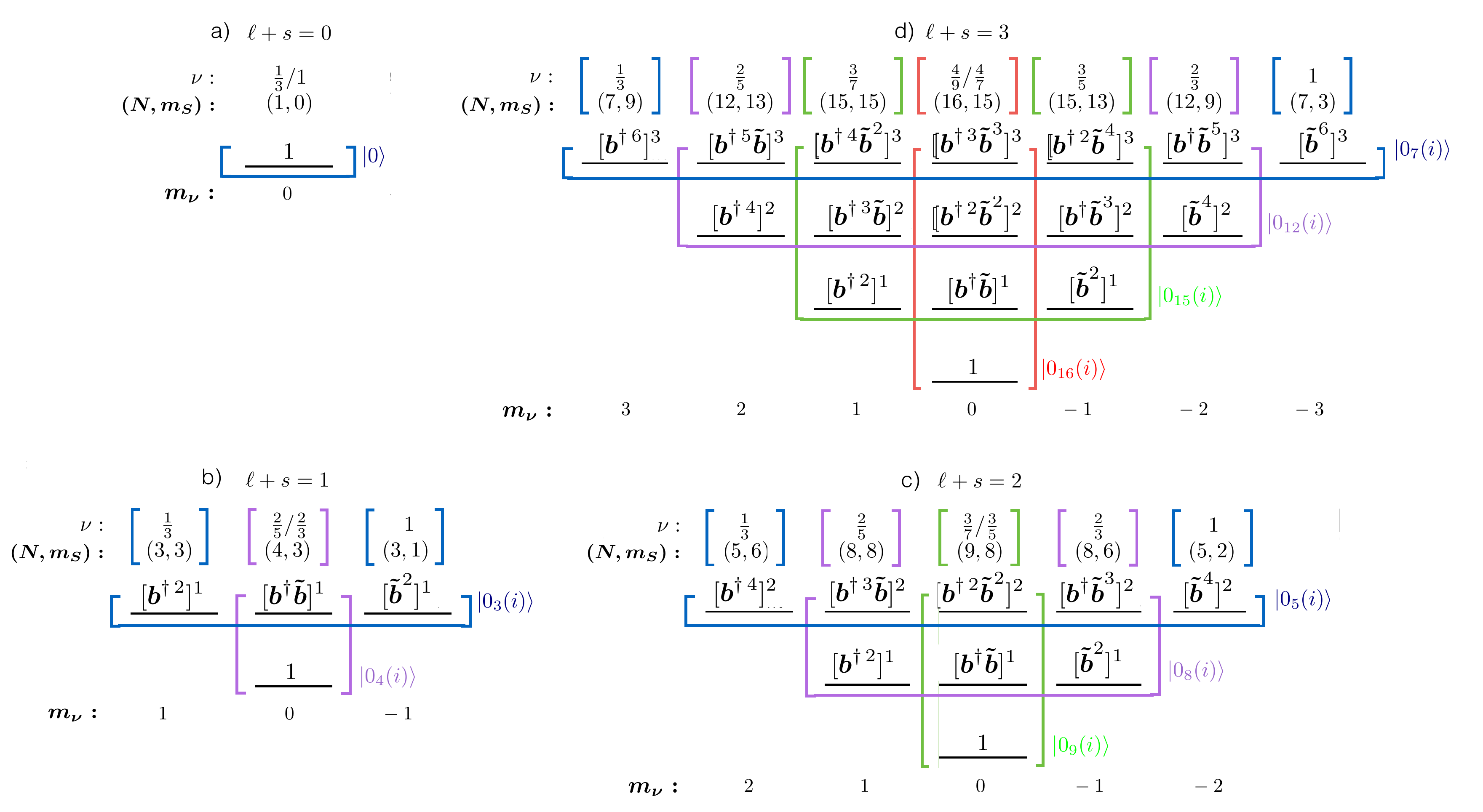}
\caption{The mirror symmetric CF multiplets for
$\ell+s$=0,1,2,3 are shown in  a), b), c), d). 
In d), for example,
four $\nu$-spin multiplets can be formed, corresponding to $p$=1 (blue), $p$=2 (purple), $p$=3 (green), and 
$p$=4 (red).   FQHE
states (ground states) are the multiplet states of maximum $|m_\nu|$, with
fillings $\nu=p/(2p+1)$ and $p/(2p-1)$.
 These patterns can be extended to arbitrary $N$.}
\label{Fig:Even}
\end{figure*}  

  \begin{figure*}[ht]   
\centering
\includegraphics[width=1.0\textwidth]{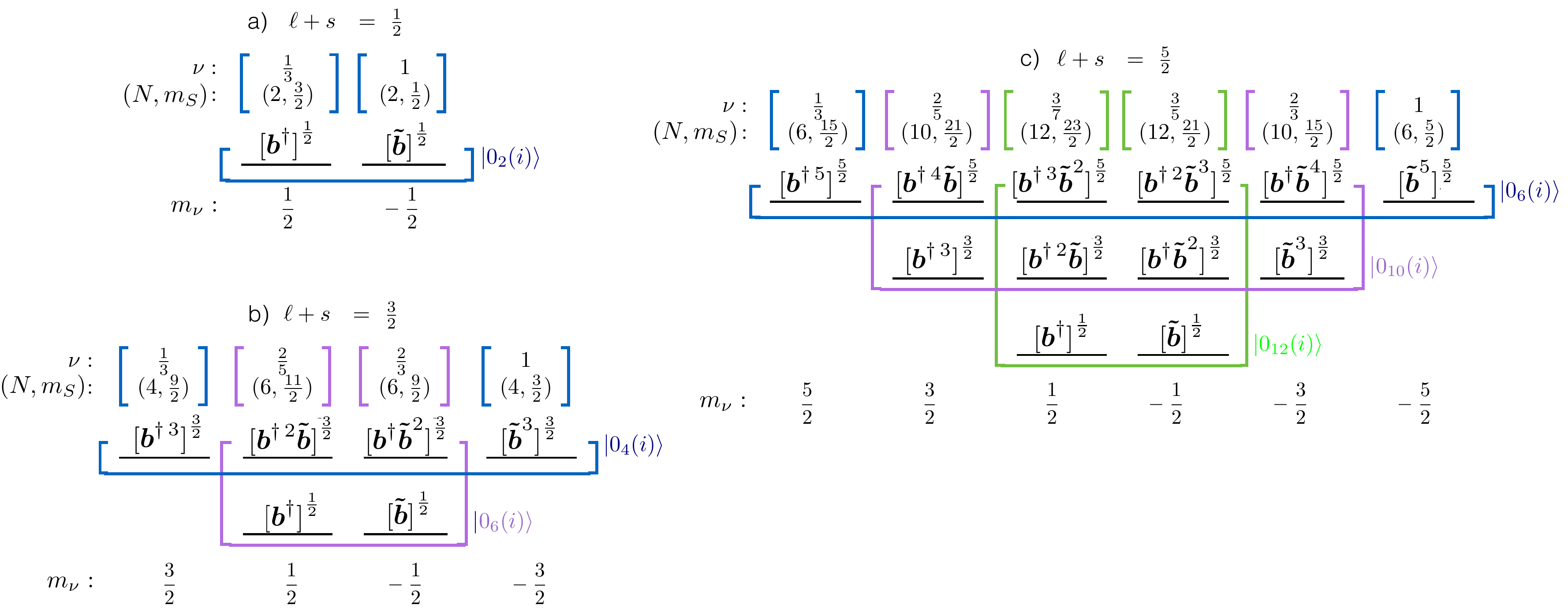}
\caption{As in Fig. \ref{Fig:Even}, but for the cases of half-integer $\ell+s$.   If Figs. \ref{Fig:Even}
and \ref{Fig:Odd} are extended to arbitrary $N$, the multiplet states of maximum $|m_\nu|$ will account for all
first-hierarchy (or $m$=3) FQHE states.}
\label{Fig:Odd}
\end{figure*} 

We construct the many-body states connected with the mirror symmetry by consider groups
of states of fixed $\ell+s$, which from Eq. (\ref{eq:relations}), implies fixed $N$.   Such states
correspond to horizontal lines across the Hilbert spaces depicted by the panels of Fig. \ref{Fig:sl},
indexed by $m_\nu$, linking subshells characterize by a fixed angular momentum.  

  \subsubsection{Electron and vortex angular momenta}
 In the lowest subshell the CF's electron and vortex angular momenta are antialigned.
 We see from Eq. (\ref{GH2A}) that the angle between these vectors gradually opens  
 with successive subshells, increasing the CF's angular momentum.   The angular momentum is generated
 by destroying energetically favored scalar pairs $u_i \cdot u_j$ between the electron and vortex, replacing them with less
 favored symmetric pairs
 $[u_i \otimes u_j]^1_{m_L}$.  Alternatively, for
 $\nu<\textstyle{1 \over 2}$ (and thus $m_\nu>0$)  this process can be viewed as successive re-alignments of vortex quanta with
 the electron spinor, as
  \begin{eqnarray}
 \Psi_{m_\nu \, m_L}^{N \, L \,\mathcal{I}=2}(i) &\sim&\sum\limits_{j \ne i}\big[ [u_i^{N-1+2m_\nu}u_j]^{{N \over 2}+m_\nu} \nonumber  \\
&&~  \otimes [u_1\cdots u_N;\bar{u}_i \bar{u}_j]^{N -2\over 2} \big]^{L=m_\nu+1}_{m_L}  \nonumber \\[2pt]
 \Psi_{m_\nu \, m_L}^{N \, L \,\mathcal{I}=3}(i) &\sim& \sum\limits_{j \ne i,k \ne i,j \ne k} \big[ [u_i^{N-1+2m_\nu}u_j u_k]^{{N+1 \over 2}+m_\nu}  \nonumber \\
&&~  \otimes [u_1\cdots u_N;\bar{u}_i \bar{u}_j \bar{u}_k ]^{N -3\over 2} \big]^{L=m_\nu+2}_{m_L} 
  \label{eq:Gamma}
 \end{eqnarray}
 and so on for additional values of $\mathcal{I}$. 
 The gap between CF shells $\mathcal{I}$+1 and $\mathcal{I}$ thus
 corresponds to the energy cost per CF of one such replacement.
 The fact that the breaking of antisymmetric pairs generates additional CF angular momentum has an obvious connection
 form of the angular momentum operator of Eq. (\ref{eq:angmom}). 
    
\subsubsection{CFs as operators with respect to the half-filled shell}
The above description of CFs is relative to the electron vacuum.  Alternatively, CFs can be expressed in terms of
valence electron creation and annihilation operators, acting on the scalar half-filled shell, which then can be regarded as
a new CF vacuum.
These valence operators -- a representation of the GH operators -- carry the CF angular momentum and 
$\nu$-spin quantum numbers $L,m_L,m_\nu$.
 One obtains
 \begin{eqnarray}
 \Psi^{N \, L}_{ m_\nu \, m_L}(i)  &=&[\mathrm{GH}]^L_{m_\nu \, m_L} |0_N(i) \rangle \nonumber \\
  &=& \left[ \boldsymbol{b}^\dagger(i)^{L+m_\nu} \boldsymbol{\tilde{b}}(i)^{L-m_\nu} \right]^{L}_{m_L} |0_N(i) \rangle 
 \label{eq:simple}
 \end{eqnarray}
 The operators creating and annihilating quanta are aligned, while
 \[ \langle \theta,\phi |0_N(i) \rangle = [u_i]^{N-1 \over 2} \odot [u_i \cdots u_N;\bar{u}_i]^{N-1 \over 2},\]
 is the CF vacuum state with $L$=0, $m_L=0$, $m_\nu=0$.  The electron-vortex symmetry at $m_\nu=0$ is thus broken by creating an excess
 or deficit of electron quanta, providing another way to think about the $\nu$-spin mirror.
 Because the creation and annihilation operators are coupled to maximum $L$, no contractions among them are allowed.
 
  A state can be turned into its  $\nu$-spin mirror by a simple conjugation of the valence operators. The mirror symmetric partners are
  \begin{eqnarray}
  \left[ \boldsymbol{b}^\dagger(i)^{L+m_\nu} \boldsymbol{\tilde{b}}(i)^{L-m_\nu} \right]^{L}_{m_L} |0_N(i) \rangle ~\longleftrightarrow~ \nonumber \\
  \left[ \boldsymbol{b}^\dagger(i)^{L-m_\nu} \boldsymbol{\tilde{b}}(i)^{L+m_\nu} \right]^{L}_{m_L} |0_N(i) \rangle 
  \label{eq:key}
  \end{eqnarray}
  Thus the mirror symmetry $m_\nu \leftrightarrow -m_\nu$ is manifested as a symmetry under valence operator conjugation
  that relates states of distinct filling,
  \[ \nu={p \over 2p+1} \leftrightarrow \bar{\nu}={p \over 2p-1}~~~~\Leftrightarrow~~~~\boldsymbol{b}^\dagger \leftrightarrow \boldsymbol{\tilde{b}} \]
  This symmetry connects states of the same $N$ and shell structures, that reside in distinct magnetic fields
   $m_S=N-1 \pm m_\nu$. 
   
   For a multiplet with $p$ subshells, $N=p(2 L_{\mathcal{I}=1} +p)$, where $L_{\mathcal{I}=1}$ is the angular momentum of the $\mathcal{I}=1$
   subshells of the conjugate $p \over 2p+1$ and $p \over 2p-1$ states.  There are $2L_{\mathcal{I}=1}+1$ members in the
   the multiplet.  This determines the $\nu={1 \over 2}$ vacua $|0_N\rangle$ in  Figs. \ref{Fig:Even} and \ref{Fig:Odd}.
  
  \subsubsection{FQHE multiplet structure}
  By labeling Figs. \ref{Fig:Even} and \ref{Fig:Odd} with the corresponding CF valence ladder
  operators, one can bring out the underlying simplicity of the multiplet structure.   For each integer (half-integer)
  value of $\ell+s$, there are $\ell+s+1$ ($\ell+s+1/2$) magnetic multiplets, 
  arranged symmetrically around $m_\nu=0$.  These multi-shelled
  multiplets are color-coded in the figures.
  Multiplet members have the same $N$: thus all members
  act on the same scalar intrinsic state $|0_N(i) \rangle$ and involve the same vortex 
  $[u_1 \cdots u_N;\bar{u}_i]^{N-1 \over 2}$.   
  
  The states of maximum $|m_\nu|$ within each multiplet
  are ground states and thus identified with FQHE states.  The lowest subshell for a FQHE state 
  has $L=|m_\nu|$, on which we can then build an angular momentum tower by filling $p$ such shells in total.  Figs. \ref{Fig:Even} 
  and \ref{Fig:Odd} illustrate another way to characterize these towers: aligned couplings of valence electron creation and annihilation
  in which factors $[\boldsymbol{b}^\dagger \otimes \boldsymbol{\tilde{b}}]^1$ are successively added.  This is the valence representation
  of the operator that acts on $u(i) \cdot u(j)$ to produce $[u(i) \otimes u(j)]^1$.     
  
  States within a multiplet that are not of maximum $|m_\nu|$ are have unoccupied lower subshells:
  these are legitimate states, translationally invariant and uniform in density, but they are clearly highly excited states, not FQHE states.
 
  If we were to continue the construction begun in  Fig. \ref{Fig:Even}, half of all the $m=3$
  FQHE states would be generated, e.g., the odd-$N$ $\nu=1/3$ and $\nu=1$ states.  One can readily see that the other
  half would be generated by the half-integer $\ell+s$ cases, illustrated in Fig. \ref{Fig:Odd}.  
  
\begin{figure}[ht]   
\centering
\includegraphics[width=0.47\textwidth]{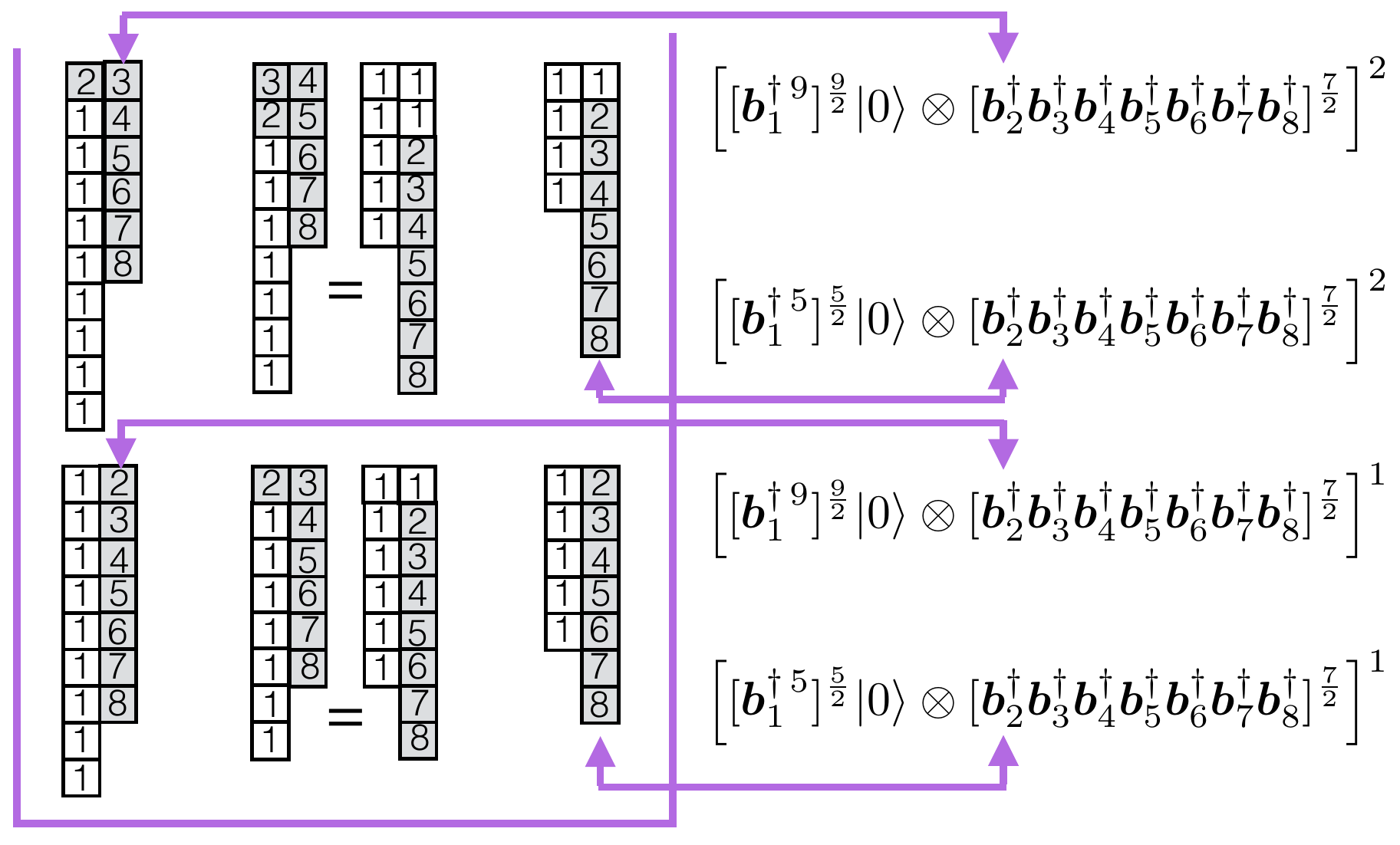}
\caption{FQHE multiplets described relative to the true  vacuum $|0\rangle$, to show by contrast the economy of the 
the valence ladder operator representation.   This example corresponds to the purple multiplet of panel c), Fig. \ref{Fig:Even}.
This CF depiction follows Eq. (\ref{eq:Gamma}), with symmetrization over the
ladder operators comprising the vortex understood.}
\label{fig:CFFinal}
\end{figure}

  Figure \ref{fig:CFFinal} is included to illustrate the less economical description of multiplet states that results when
  CFs are expressed relative to the vacuum, rather than the half-filled shell.   The cases illustrated correspond
  to the three multiplets of panel c), Fig. \ref{Fig:Even}.   
  
\subsection{Electron-vortex algebraic structure}
The multiplets just illustrated reflect an underlying electron-vortex algebraic structure.
Our CFs are constructed as good spherical tensors, and thus are eigenstates of the 
total angular momentum and $L_z$.  For the CF labeled 1, the operator
components are
\begin{equation}
\hat{L}_{1m_L} ={\hbar \over \sqrt{2}} \Big( [\boldsymbol{b}^\dagger(1) \otimes \boldsymbol{\tilde{b}}(1)]_{1m_L} + \sum_{j=2}^N  [\boldsymbol{b}^\dagger(j) \otimes \boldsymbol{\tilde{b}}(j)]_{1m_L} \Big)
\end{equation}

PH operations of interest involve adding or removing a quantum from a vortex.  Addition could be done, for example,
by an aligned coupling of $\boldsymbol{b}^\dagger(N)$ to the vortex $[\boldsymbol{b}^\dagger(1) \cdots \boldsymbol{b}^\dagger(N-1)]^{N-1 \over 2}$.  But it is more attractive at this point to
treat the vortex components more symmetrically.  We
introducing the vortex creation and annihilation operators for CF 1,
\begin{eqnarray}
[\boldsymbol{v}^\dagger]^{1 \over 2}_{m_L} (1)&=&P_N \,\textstyle\sqrt{{N(N-1) \over 2}} \Big[ [\boldsymbol{b}^\dagger(2) \cdots \boldsymbol{b}^\dagger(N) ]^{{N-1 \over 2}}  \nonumber \\
&&~~~~~~ \otimes~ [\boldsymbol{\tilde{b}}(2) \cdots \boldsymbol{\tilde{b}}(N-1)]^{N-2 \over 2} \Big]_{{1 \over 2} m_L} \nonumber \\
~[\boldsymbol{\tilde{v}}]^{1 \over 2}_{m_L}(1)&=&\textstyle\sqrt{{N(N-1) \over 2}} \Big[ [\boldsymbol{b}^\dagger(2) \cdots \boldsymbol{b}^\dagger(N-1) ]^{{N-2 \over 2}}  \nonumber \\
&&~~~~~~ \otimes~ [\boldsymbol{\tilde{b}}(2) \cdots \boldsymbol{\tilde{b}}(N)]^{N-1 \over 2} \Big]_{{1 \over 2} m_L} \, P_N. 
\end{eqnarray}
We have included explicit projection operators $P_N$, indicating vortices are produced or destroyed corresponding to a system with $N$ total
electrons.  In using these operators, it is understood that $N$ is chosen so the vortex in the ket is annihilated by the term on right.
With this rule one finds
\begin{equation}
 ~[ \boldsymbol{\tilde{v}}_{m_L}, \boldsymbol{v}^\dagger_{{m_L}^\prime} ] = \delta_{m_L,-m_L^\prime} (-1)^{{1 \over 2} +m_L} 
\end{equation}
Just as $\boldsymbol{b}^\dagger \cdot \boldsymbol{\tilde{b}}$ is the number operator for the CF's electron spinor, counting the number
of quanta, 
$\boldsymbol{v}^\dagger \cdot \boldsymbol{\tilde{v}}$ plays the same role for the vortex
\begin{eqnarray}
&& \boldsymbol{v}^\dagger \cdot \boldsymbol{\tilde{v}} ~[\boldsymbol{b}^\dagger(2) \cdots \boldsymbol{b}^\dagger(N) ]^{N-1 \over 2}_{m_L} \nonumber \\  
&&~~~~~~~=(N-1)[\boldsymbol{b}^\dagger(2) \cdots \boldsymbol{b}^\dagger(N)]^{N-1 \over 2}_{m_L} . 
\end{eqnarray}
The CF angular momentum operators then take on a more symmetric electron-vortex form
\begin{flalign}
&\hat{L}_{1m_L}(1) \equiv \hat{L}^e_{1m_L}(1) +  \hat{L}^v_{1m_L}(1) \nonumber \\
&~~~= {\hbar \over \sqrt{2}} \Big( [\boldsymbol{b}^\dagger(1) \otimes \boldsymbol{\tilde{b}}(1)]_{1m_L} +  [\boldsymbol{v}^\dagger(1) \otimes \boldsymbol{\tilde{v}}(1)]_{1m_L} \Big) 
\end{flalign}

Using the noninteracting, multi-LL case as a model,
we define the four-component creation operator for CF 1
\begin{equation}
\boldsymbol{d}^\dagger_{m_\nu \, m_L}(1) \equiv \left( \begin{array}{c} \boldsymbol{d}^\dagger_{~{1 \over 2}~{1 \over 2}}(1) \\ \boldsymbol{d}^\dagger_{~{1 \over 2}~-{1 \over 2}}(1) \\ \boldsymbol{d}^\dagger_{-{1 \over 2}~{1 \over 2}}(1) \\ \boldsymbol{d}^\dagger_{-{1 \over 2}~-{1 \over 2}}(1) \end{array} \right)
\equiv  \left( \begin{array}{c} \boldsymbol{b}^\dagger_{~{1 \over 2}}(1) \\ \boldsymbol{b}^\dagger_{-{1 \over 2}}(1) \\ \boldsymbol{v}^\dagger_{~{1 \over 2}}(1) \\ \boldsymbol{v}^\dagger_{-{1 \over 2}}(1) \end{array} \right)
\end{equation}
and similarly
\begin{equation}
\boldsymbol{\tilde{d}}_{m_\nu \, m_L}(1) \equiv \left( \begin{array}{c} \boldsymbol{\tilde{d}}_{~{1 \over 2}~{1 \over 2}}(1) \\ \boldsymbol{\tilde{d}}_{~{1 \over 2},~-{1 \over 2}}(1) \\ \boldsymbol{\tilde{d}}_{-{1 \over 2},~{1 \over 2}}(1) \\ \boldsymbol{\tilde{d}}_{-{1 \over 2},-{1 \over 2}}(1) \end{array} \right)
\equiv  \left( \begin{array}{c} -\boldsymbol{\tilde{v}}_{~{1 \over 2}}(1) \\ -\boldsymbol{\tilde{v}}_{-{1 \over 2}}(1) \\ \boldsymbol{\tilde{b}}_{~{1 \over 2}}(1) \\ \boldsymbol{\tilde{b}}_{-{1 \over 2}}(1) \end{array} \right)
\end{equation}
We find the commutator
\begin{equation}
[\boldsymbol{\tilde{d}}_{m_\nu,m_L},\boldsymbol{d}^\dagger_{m_\nu^\prime,m_L^\prime} ] = \delta_{m_\nu,-m_\nu^\prime} \delta_{m_L,-m_L^\prime} (-1)^{{1 \over 2}-m_\nu+{1 \over 2}-m_L}
\end{equation}

We can build associated operators.  $\hat{L}_{1m_L}$ is identified as the bilinear operator
carrying $\{S^\nu,m_\nu;L,m_L\}=\{0,0;1,m_L\}$
\[ \hat{L}_{1m_L} ={\hbar} \left[ \boldsymbol{d}^\dagger(1) \otimes \boldsymbol{\tilde{d}}(1) \right]_{S^\nu=0 \, m_\nu=0; L=1 \,m_L} \]
with the CFs satisfying
 \begin{eqnarray}
\hat{L}_0 \,  \Psi^{N \, L}_{m_\nu \, m_L}(1)  &=& \hbar m_L \, \Psi^{N \, L}_{m_\nu \, m_L}(1)\nonumber \\
\hat{L}_+ \Psi^{N \, L}_{m_\nu \, m_L}(1) &=&\hbar  \sqrt{(L-m_L)(L+m_L+1)}\, \Psi^{N \, L}_{m_\nu \, m_L+1}(1)   \nonumber \\
\hat{L}_-  \Psi^{N \, L}_{m_\nu \, m_L}(1) &=& \hbar  \sqrt{(L+m_L)(L-m_L+1)} \, \Psi^{N \, L}_{m_\nu \, m_L-1}(1) ~~~~~
\end{eqnarray}

Similarly we can form the $\nu$-spin operator triad, analogous to $\hat{S}$ of the noninteracting problem,
\begin{equation}
 \hat{S}^\nu_{1m_\nu} = \left[ \boldsymbol{d}^\dagger(1) \otimes \boldsymbol{\tilde{d}}(1) \right]_{S^\nu=1 \,m_\nu; L=0 \, m_L=0} 
 \end{equation}
 These are angular momentum scalars, but
 rank-one tensors in $\nu$-space,
\begin{eqnarray}
\hat{S}_0^\nu(1) &=& {1 \over 2} \left[ \boldsymbol{b}^\dagger(1) \odot \boldsymbol{\tilde{b}}(1) -\boldsymbol{v}^\dagger(1) \odot \boldsymbol{\tilde{v}}(1) \right] \nonumber \\
\hat{S}^\nu_+(1) &\equiv& -\sqrt{2} \hat{S}^\nu_{11}=\hat{S}^\nu_x+i \hat{S}^\nu_y = \boldsymbol{b}^\dagger(1) \odot  \boldsymbol{\tilde{v}}(1)  \nonumber \\
\hat{S}^\nu_-(1) &\equiv& \sqrt{2} \hat{S}^\nu_{1-1} = \hat{S}^\nu_x-i \hat{S}^\nu_y =  \boldsymbol{{v}}^\dagger(1)  \odot \boldsymbol{\tilde{b}} (1)
\label{eq:raising}
\end{eqnarray}
The implicit dependence of $\boldsymbol{v}^\dagger$ and $\boldsymbol{\tilde{v}}$ on $N$ is again defined by
the rule that these operators first act to annihilate the vortex in the ket.  

$\hat{S}^\nu_0$ is the operator associated with the CF's second magnetic index,
\begin{equation}
\hat{S}^\nu_0 \,  \Psi^{N \, L}_{m_\nu \, m_L}(1)  = m_\nu \, \Psi^{N \, L}_{m_\nu \, m_L}(1).
\end{equation}
Thus we have identified the complete set of commuting operators for the GH$^2$ CFs.
As $\hat{S}_\nu$ is an angular momentum scalar,
$ [\hat{S}_0^\nu,\hat{L}_z]=[\hat{S}_0^\nu,\hat{L}^2]=0$,
allowing us to use $L, \,m_L, \, m_\nu$
as simultaneous CF quantum labels (along with $N$).
It is also straightforward to show
\begin{equation}
\hat{S}^{v \,2}  \Psi^{N \, L}_{m_\nu \, m_L}= \hat{L}^2 \Psi^{N \, L}_{m_\nu \, m_L}=L(L+1)  \Psi^{N \, L}_{m_\nu \, m_L}
\end{equation}
accounting for the association of both magnetic indices with the CF's total angular momentum.  

The raising/lowering operators $S_+^\nu$ and $S_-^\nu$ transfer a quantum from/to the vortex 
to/from the electron spinor.  They play a crucial role in the second (conventional) PH symmetry,
described in the next section: they generate algebraic connections between states of different electron number
residing in the same magnetic field (in contrast to our $\nu$-spin symmetry that relates states
residing in different magnetic fields but having the same $N$).    Because they link states of 
different $m_\nu$, the relative normalization of CFs must be specified.  We adopt the
following normalization (consistent with that for electron spinors)
\begin{flalign}
&\tilde{\Psi}^{N \, L}_{m_\nu \, m_L }(1) = {1  \over \sqrt{(N-1+2m_\nu)!}}~~~~~~~~~~~~~~~~~~~~~~~~~~~~~\nonumber \\
&\times  [[\boldsymbol{b}^\dagger(1)]^{{N-1 \over 2}+m_\nu} \otimes [\boldsymbol{b}^\dagger(2) \cdots \boldsymbol{b}^\dagger(N)]^{{N-1 \over 2}-m_\nu} ]_{m_L}^L|0\rangle 
\label{eq:norm}
\end{flalign}
denoting normalized CFs by $\tilde{\Psi}$.  Recall the magnetic field strength $m_S=N-1+m_\nu$.  We find 
\begin{eqnarray}
\hat{S}_+^\nu \, \tilde{\Psi}^{N \, L}_{m_\nu  \, m_L} &=& \sqrt{(L+m_\nu+1)(L-m_\nu)}  \, \tilde{\Psi}^{N-1 \, L}_{m_\nu+1  \, m_L} \nonumber \\
\hat{S}_-^\nu \, \tilde{\Psi}^{N \, L}_{m_\nu  \, m_L} &=& \sqrt{(L-m_\nu+1)(L+m_\nu)}  \, \tilde{\Psi}^{N+1 \, L}_{m_\nu-1  \, m_L} ~~~~
\end{eqnarray}
These operations preserve $m_S$. From $\hat{S}_+^\nu$ and $\hat{S}_-^\nu$  one can 
construct operators diagonal in $m_\nu$
\begin{eqnarray} \
\hat{S}^\nu_+\hat{S}^\nu_- \, \tilde{\Psi}^{N \, L}_{m_\nu \, m_L} &=&(L+m_\nu)(L-m_\nu+1) \, \tilde{\Psi}^{N \, L}_{m_\nu \, m_L} \nonumber \\
\hat{S}^\nu_-\hat{S}_+ \, \tilde{\Psi}^{N \, L}_{m_\nu \, m_L} &=&(L-m_\nu)(L+m_\nu+1) \, \tilde{\Psi}^{N \, L}_{m_\nu \, m_L}~~~
\label{eq:s2}
\end{eqnarray}
which are related to $\hat{S}_0^\nu$ by
\begin{equation}
\hat{S}^\nu_+ \hat{S}^\nu_- - \hat{S}^\nu_- \hat{S}^\nu_+ = 2 \hat{S}_0^\nu.
\label{eq:comm}
\end{equation}
These relations are important to CF Hamiltonians we discuss later.

The analog of $\nu$-spin familiar in QCD, isospin, is used as a quantum label, despite electromagnetic
interactions that break isospin at a few tenths of a percent in the nucleon, and at higher levels in nuclei.
The numerical studies with GH$^2$ CFs in \cite{GH2} indicate that $\nu$-spin symmetry for modest $N$ holds in the FQHE
to about 0.1\%, suggesting $\nu$-spin may be at least as good a quantum number as isospin.
In the QCD case the algebraic relationship among states implies approximate degeneracies: e.g.,
the masses of the $^1S_0$ states of the two-nucleon system $nn$, $np$, and $pp$ 
with isospin $T=1$ and $m_\tau=-1,0,1$, are equal to within about 0.3\%. Similarly, we will argue later in this
paper that $\nu$-spin symmetry implies degeneracies between mirror CFs.

\section{Particle-Hole Symmetry}
We now turn to a second symmetry, PH conjugation.  
FLL wave functions computed numerically with the inclusion of all electron degrees of freedom would, of course, exhibit exact PH symmetry.
Our goal here is not to construct a CF representation exhibiting a similar exact symmetry  (although this appears to us possible,
as we have explicit representations for CFs at both $\nu<\textstyle{1 \over 2}$ and
$\nu>\textstyle{1 \over 2}$ that can be combined symmetrically).  Rather, we are interested
in the algebraic manifestation of PH symmetry when wave functions are expressed in their CF forms.  The manifestion
will turn out to involve the behavior of CFs under electron-vortex exchange.

%Our discussion proceeds through the following steps:
%\begin{enumerate}
%\item We discuss the relationship between PH conjugate many-electron states in the context of the
%$\nu$-spin multiplets of the previous section.
%\item Unlike the case of the previous section where $\nu$-spin symmetry under the constraint of constant $N$
%allowed us to discuss CF and many-electron states together, for PH symmetry and fixed $S$, it is more
%convenient to first discuss the multiplet structure of single CFs.  The needed multiplets of fixed $S$ are generated by
%families of valence P(electron)-H(vortex) excitations.
%\item We then argue that it is easiest to visualize the associated many-electron states relevant to PH
%symmetry in terms of nearly exact CF decompositions of the filled shell, indexed by $\bar{N}-N$.  This picture
%of PH symmetry seems to us to differ from that purposed by Son.
%\item We then turn to the Hamiltonians that the CFs satisfy.  We show that there are minimal Pauli (quadratic) and Dirac (linear)
%forms for the Hamiltonian at $\nu=\textstyle{1 \over 2}$ that can be identified by their correspondence to the known IQHE case.
%The form can be easily generalized to all fillings.
%\item We discuss the possibility that the physics of the full (as opposed to minimal) Hamiltonian can be
%related to the Coulomb Hamiltonian for the IQHE, reflecting the scale invariance of the Coulomb interaction.
%\end{enumerate}

\begin{figure*}[ht!]   
\centering
\includegraphics[width=1.0\textwidth]{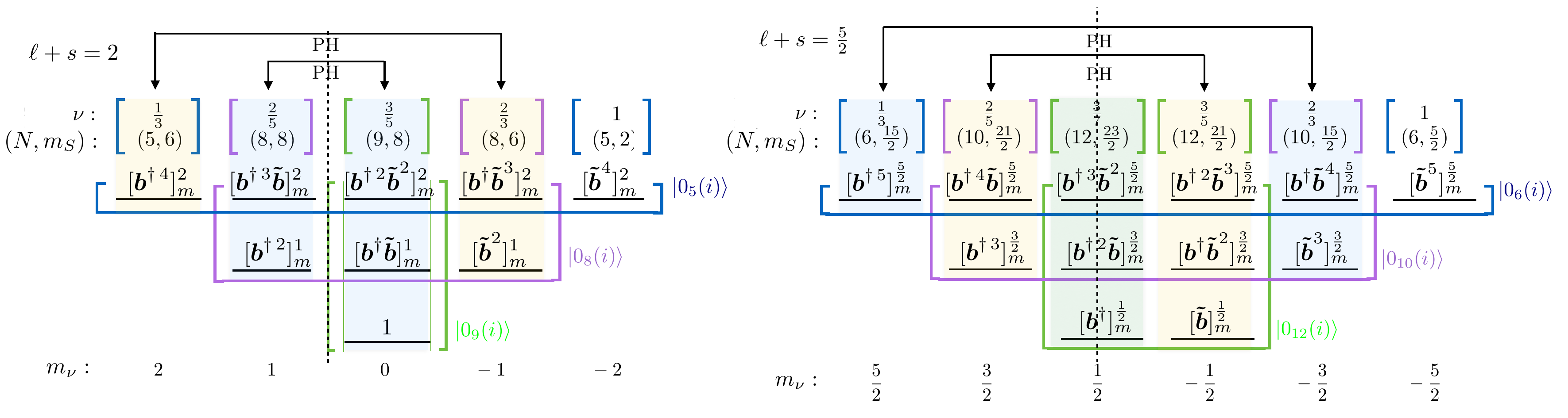}
\caption{FQHE PH conjugate pairs, indicated by background shading, for two
cases from Figs. \ref{Fig:Even} and \ref{Fig:Odd}, illustrated for the constant-$N$ multiplets of Sec. IV.}
\label{Fig:SymPH}
\end{figure*} 

\begin{figure*}[ht!]   
\centering
\includegraphics[width=0.9\textwidth]{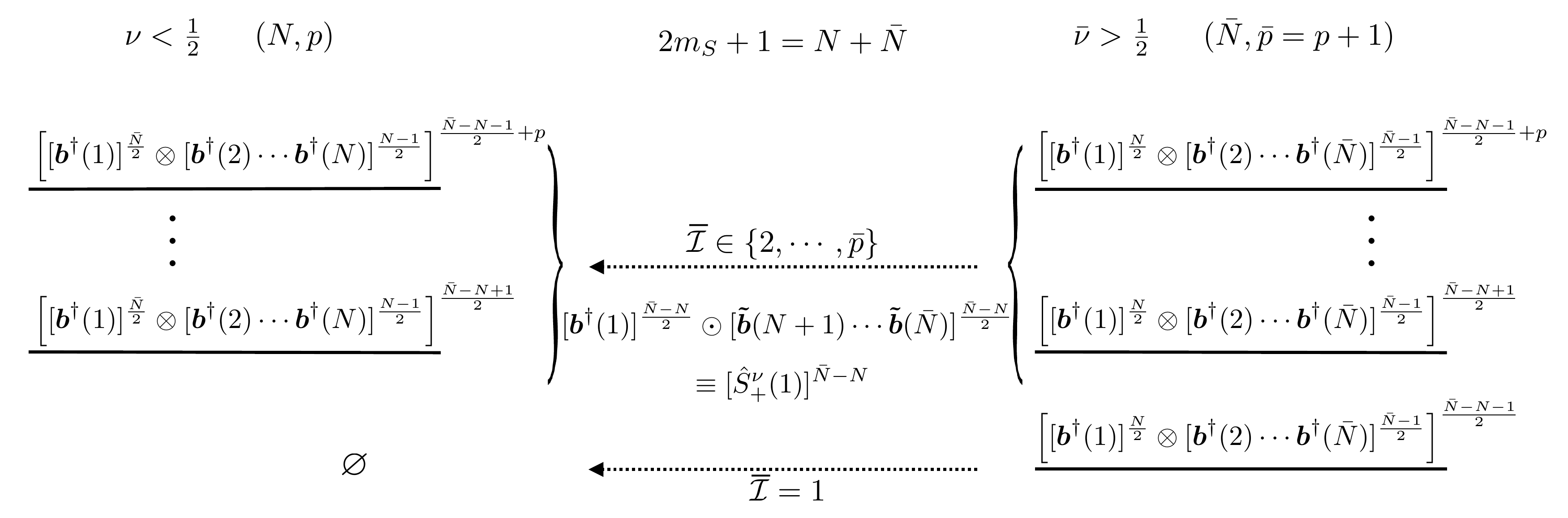}
\caption{The CF relationship for conjugate PH states.  The conjugation operator is also shown.  Note the $N \leftrightarrow \bar{N}$ reverses
in the number of quanta carried by the CF's electron and vortex components, on PH conjugation.}
\label{Fig:PHpattern}
\end{figure*} 

\subsection{PH quantum number relationships}
PH symmetry connects states of different electron number that reside in the same magnetic field:  $m_S=\bar{m}_S$, $N \ne \bar{N}$, and $N+\bar{N}=2m_S+1$ (where the bars indicate conjugate state labels).   
Although the multiplets we have previously defined correspond to fixed $N$, not fixed $m_S$, they contain
all FQHE states, and so can be used to identify the basic structure of PH conjugate states.
The CFs for a FQHE state and its conjugate must have the same number of total
quanta ($n_e+n_v=\bar{n}_e+\bar{n}_v$)
in order to belong to the same magnetic field.  
Using our expressions for the CFs and the constraint $N + \bar{N}=2m_S+1$,
the following relations among state quantum numbers are easily obtained
\begin{eqnarray}
N-1+m_\nu =m_S &=&  \bar{N} -1 + \bar{m}_\nu \nonumber \\
 1 &=&   m_\nu +\bar{m}_\nu\nonumber \\
 \bar{N}&=& N+2 m_\nu -1  \nonumber \\
 \bar{p}&=&p+1 \nonumber \\
 (\bar{\ell},\bar{s})&=&(s+{1 \over 2},\ell-{1 \over 2}) \nonumber \\
 \bar{\ell}+\bar{s} &=& \ell+s
\end{eqnarray}
From these relationships one sees $\ell+s$ is unchanged, while the difference in $N$ and $\bar{N}$ depends only on
$m_\nu$: this suggests an underlying $\nu$-spin multiplet structure.
The multiplet partners linked by PH symmetry are not mirror, as
$m_\nu+\bar{m}_\nu=1$, not 0.  The constraint  $\bar{p}=p$+1 tells us that PH symmetry connects states 
that differ by one in their subshell numbers.

In Fig. \ref{Fig:SymPH} panels from  Figs. \ref{Fig:Even} and \ref{Fig:Odd} have been redrawn to illustrate the positioning 
of PH conjugate states in our fixed $N$ $\nu$-spin multiplets. The cases for integer and half-integer angular momentum are 
similar.  In Figure \ref{Fig:SymPH} every $\nu<{1 \over 2}$ state labeled by $N$, $p$, and $m_\nu$ has a PH conjugate
state living in the same magnetic field (same $S$) labeled by $\bar{N}$, $\bar{p}=p+1$, and $\bar{m}_\nu = 1-m_\nu$.
The figures illustrate the offset -- ${m_\nu + \bar{m}_\nu \over 2}={1 \over 2}$ -- and shell asymmetry -- $\bar{p}=p+1$ --
noted above.  The $\mathcal{I}=1$ subshell of the $\bar{\nu}$ state is the subshell without a partner: this
subshell contains $\bar{N}-N$ CFs.

It is important to note that in these figures, the conjugate states belong to different multiplets: the GH operators that 
generate these states act on different vacua $|0_N \rangle$ and $|0_{\bar{N}} \rangle$.  Thus the CFs occupying $p$
subshells of the $\nu<{1 \over 2}$ state do not evolve into their PH conjugate partners as $m_\nu \rightarrow \bar{m}_\nu$.
These multiplets have fixed $N$, preventing changes in the vortex.

Despite these various asymmetries, the relationship between PH conjugate states is very simple., as
illustrated in Fig. \ref{Fig:PHpattern}.   With the exception of the
annihilated $\bar{p}=1$ subshell that accounts for the number difference $\bar{N}-N$, we see
that PH symmetry is manifested microscopically as a simple exchange in the quanta, $N \leftrightarrow \bar{N}$,
carried by the electron and vortex components of the conjugate CFs,
\begin{eqnarray}
\tilde{\Psi}^{N \, L \, \nu<{1 \over 2}}_{m_\nu \, m_L }(1) =~~~~~~~~~~~~~~~~~~~~~~~~~~~~~~~~~~~~~~~~~~~~~ \nonumber \\
\textstyle{{1 \over \sqrt{\bar{N!}}}} \big[ \left[\boldsymbol{b}^\dagger(1) \right]^{\bar{N} \over 2} \otimes \left[ \boldsymbol{b}^\dagger(2) \cdots \boldsymbol{b}^\dagger(N) \right]^{N-1 \over 2} \big]_{m_L}^L |0 \rangle~~~ \nonumber \\
\tilde{\Psi}^{\bar{N} \, L \, \nu>{1 \over 2}}_{\bar{m}_\nu \, m_L}(1) =~~~~~~~~~~~~~~~~~~~~~~~~~~~~~~~~~~~~~~~~~~~~~  \nonumber \\
\textstyle{{1 \over \sqrt{N}!} }\big[ \left[\boldsymbol{b}^\dagger(1) \right]^{{N \over 2}} \otimes \left[ \boldsymbol{b}^\dagger(2) \cdots \boldsymbol{b}^\dagger(\bar{N}) \right]^{\bar{N}-1 \over 2} \big]_{m_L}^L |0 \rangle ~~~
\label{eq:NNbar}
\end{eqnarray}
where $L$ takes on ascending values beginning with $|m_\nu|$, and $N+\bar{N}=2m_S+1$.  

\subsection{PH conjugation and $\nu$-spin}
In analogy with  $\nu$-spin symmetry, one can identify an operator that converts the CFs for the
$\bar{\nu}$ state into those of the conjugate $\nu$ state.   This operator is a angular momentum scalar,
carries $m_\nu=\bar{N}-N$ and is fully aligned in $\nu$-spin, and preserves the magnetic field and thus the sum of CF 
electron and votex quanta, 
\begin{eqnarray}
&&\hat{\Lambda}_+[i;N+1,\cdots,\bar{N}]  \nonumber \\
&&~~~~~~~\equiv  \left[\hat{S}^\nu_+(i)\right]^{\bar{N}-N} = \left[ \boldsymbol{b}^\dagger(i) \cdot \boldsymbol{\tilde{v}}(i)\right]^{\bar{N}-N}  \nonumber \\
&&~~~~~~~ = [\boldsymbol{b}^\dagger(i)]^{\bar{N}-N \over 2} \odot [\boldsymbol{\tilde{b}}(N+1) \cdots \boldsymbol{\tilde{b}}(\bar{N})]^{\bar{N}-N \over 2} ~~~
\end{eqnarray}
On acting on a CF it yields
\begin{eqnarray}
&& \hat{\Lambda}_+[i;N+1,\cdots,\bar{N}]~\tilde{\Psi}^{\bar{N} \, L}_{\bar{m}_\nu \, m_L}(i)  = \nonumber \\
&& \left\{ \begin{array}{ll}  \left[ {(L+m_\nu)! (L-1+m_\nu)! \over (L-m_\nu)! (L+1-m_\nu)! } \right]^{1 \over 2} \tilde{\Psi}^{N \, L}_{m_\nu \, m_L}(i) & ~~~L>|\bar{m}_\nu| \\[2ex] 0 & ~~~L=|\bar{m}_\nu| \end{array} \right.
\label{opPH}
\end{eqnarray}
where the allowed values of $\bar{m}_\nu$ are $0, \,-\textstyle{1 \over 2},\, -\textstyle{1}, \dots$ with $m_\nu=1-\bar{m}_\nu$.

One can decompose $\hat{\Lambda}_+$ into a succession of  $\bar{N}-N$ operations $\hat{S}^\nu_+$ that evolve the CF belonging to
the $\nu>{1 \over 2}$ state into the CF for the PH conjugate $\nu<{1 \over 2}$ state.   Thus we learn the PH symmetry connects
states through multiplets created by the $\nu$-spin raising and lowering operator $\hat{S}^\nu_\pm$.  Under $\nu$-spin
$m_S$ is held constant, but particle numbers evolve.  We thus should be able to establish the algebraic relationships between the CFs 
under PH conjugation by constructing this new set of constant-$m_S$ multiplets.   This will help us to
\begin{enumerate}
\item identify the valence operator description of the CFs appearing in these new multiplets;
\item understand the connection between PH conjugation and electron-vortex symmetry; and
\item restore the mirror symmetry that naively appears broken when we compare PH conjugate pairs.
\end{enumerate}

 \subsection{The GH$^\nu$ operators and electron-vortex symmetry}
 Here we seek a new set of operators  GH$^\nu$, analogous to the GH operators of Eq. (\ref{eq:simple}) but related to one another by
 the raising the lowering operations $\hat{S}^\nu_\pm$, that we can use to evolve CFs under the constraint of a
 constant magnetic field $m_S$, but changing $N$.  Unlike the GH case, these new operators must involve
 both electron and vortex excitations relative to the half-filled shell.
 
 Operators that are
 constructed as good spherical tensors in $L, m_\nu$ will automatically transform properly under $\hat{S}^\nu_\pm$.  The condition of 
 a fixed magnetic field requires that the operators not alter the total number of CF quanta: thus the operators must include an equal
 number of creation and annihilation operators.  We also require that the operators be valence operators, acting on a 
 vacuum state $|0_{N_{1 \over 2}} \rangle$ associate with the half-filled shell, where $m_\nu=0$.
 
 We first consider the case of integer angular momentum.  Consider
  \begin{eqnarray}
 \Psi^{N \, L}_{ m_\nu \, m_L}(i)  &=&[\mathrm{GH}^\nu]^{ L}_{m_\nu \, m_L} |0_{N_{1 \over 2}}(i) \rangle \nonumber \\
  &=& \left[ \boldsymbol{d}^\dagger(i)^{L} \otimes \boldsymbol{\tilde{d}}(i)^{L} \right]^{L}_{m_\nu \, m_L} |0_{N_{1 \over 2}}(i) \rangle 
 \label{eq:simplePH}
 \end{eqnarray}
 where all couplings are fully aligned in both angular momentum and $\nu$-spin.  The operators satisfy the conditions we
 have described.   As we are now connecting CFs of different particle number (different vortices), the fixed vacuum state 
 is defined relative to the half-filled shell:  unlike in the case of the GH operators,
 $N$ varies across the multiplet.  $N$ and $\bar{N}$ are related to the vacuum state by
 \begin{equation}
 N_{1 \over 2}=m_S+1={N+\bar{N}+1 \over 2}=
 p^2+2 L_{\mathcal{I}=1} p+L_{\mathcal{I}=1}
 \label{eq:simpleVacuum}
 \end{equation}
 where $N$ and $\bar{N}$ are the particle numbers of the PH conjugate CFs connected by the multiplet, related of course to $m_S$ by
 $2m_S+1=N + \bar{N}$. 
 As in our discussion of the GH operators, $L_{\mathcal{I}=1}$ is the angular momentum of the $\nu<{1 \over 2}$ state
 $\mathcal{I}=1$ subshell, which determines the rank of the multiplet, and $p$ is in the number of filled subshells
 in that state.  
 
 As $m_S$ is constant and the operators properly transform under $\nu$-spin, it is sufficient to demonstrate that 
 Eq. (\ref{eq:simplePH}) contains a multiplet member.  Evaluating this for $p=1$ and $m_\nu=L$ yields
   \begin{eqnarray}
 \Psi^{N=2L+1 \, L}_{ m_L}(i)
  &=& \left[ \boldsymbol{b}^\dagger(i)^{L} \otimes \boldsymbol{\tilde{v}}(i)^{L} \right]^{L}_{m_L} |0_{3L+1}(i) \rangle  \nonumber \\
    &=& \left[ \boldsymbol{b}^\dagger(i)^{2L} \right]^{L}_{m_L} |0_{2L+1}(i) \rangle
 \label{eq:simple}
 \end{eqnarray}
which we recognize as the CF for the $\nu={1 \over 3}$ Laughlin state.  As with the GH construction, this state and the
lowering algebra determines the entire hierarchy of GH$^\nu$ operators.

 \begin{figure*}[ht]   
\centering
\includegraphics[width=0.65\textwidth]{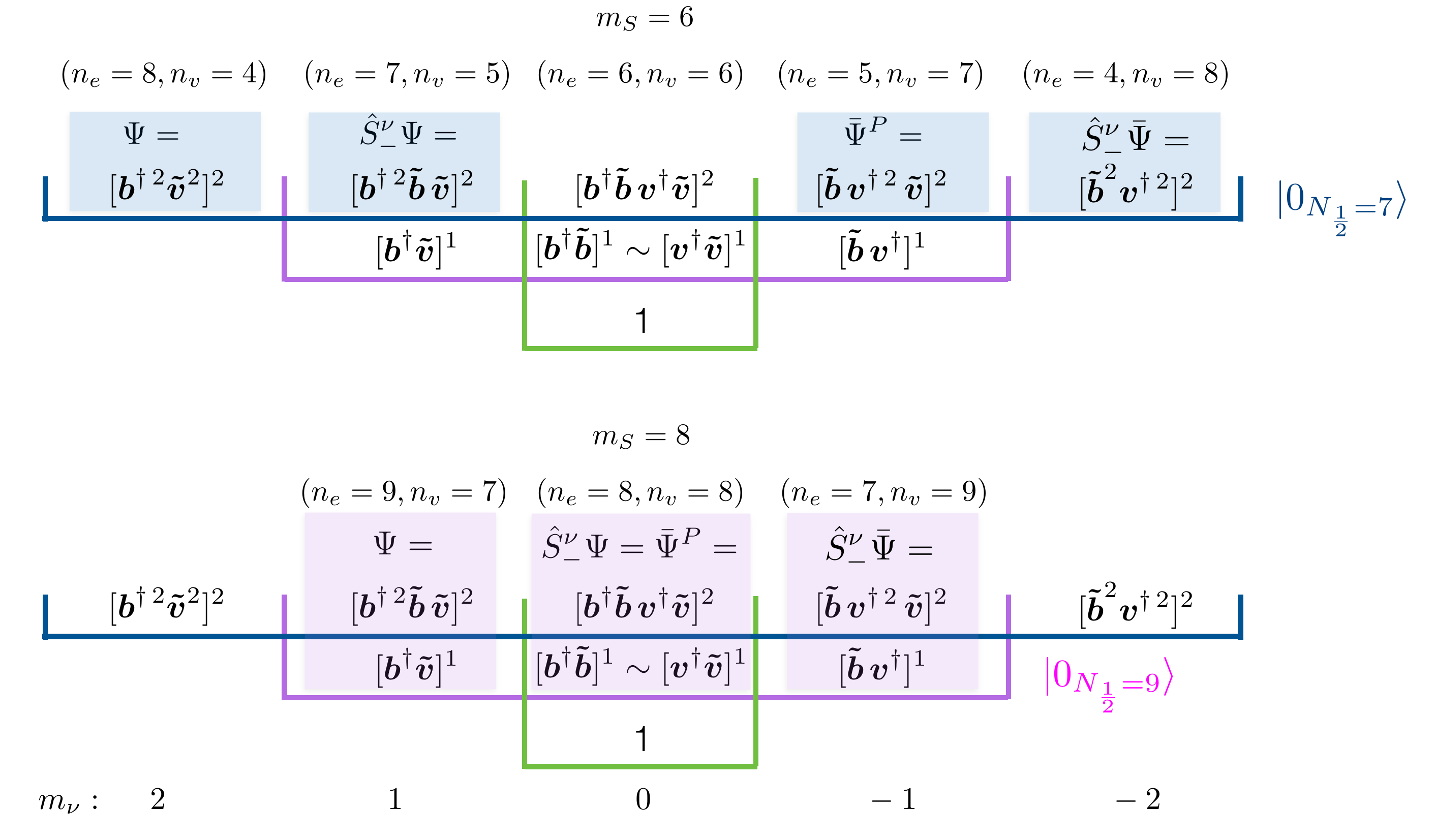}
\caption{The GH$^\nu$ valence operators that connect states linked by $\hat{S}^\nu_\pm$, illustrated for the integral angular
momentum case of Fig. \ref{Fig:PHpattern}.  PH symmetry arises in CF representations as a PH electron-vortex exchange 
symmetry, linking mirror $\nu$-spin states along a trajectory of constant magnetic field strength $m_S$.}
\label{fig:electronvortex_integral}
\end{figure*}

 The GH$^\nu$ operators are given in Fig. \ref{fig:electronvortex_integral}, which the reader should envision extended to an arbitrary number of
 rows.   (Contrast with the left panel of Fig. \ref{Fig:PHpattern}, the similar GH case where PH CFs belong to distinct multiplets.)
 We have highlighted four states in the figure,
 $\Psi$, $\hat{S}^\nu_- \Psi$, $\bar{\Psi}^P$, and $\hat{S}^\nu_- \bar{\Psi}$.   It is apparent that one can ``restore" a mirror symmetry
 between PH conjugate states by forming the pairs
 \begin{equation}
   \Psi \leftrightarrow \hat{S}^\nu_- \bar{\Psi}~~~~~~~~~\hat{S}^\nu_- \Psi \leftrightarrow \bar{\Psi}^P  
   \label{eq:partners}
   \end{equation}
 instead of focusing on $\Psi \leftrightarrow \bar{\Psi}$.   Here $\bar{\Psi}^P$ is the $\nu>{1 \over 2}$ state projected on the 
 upper $p$ subshells, $\mathcal{I}=2, \cdots, \bar{p}=p+1$.   Note the $\Psi$ can be fully reconstructed from $\hat{S}^\nu_- \bar{\Psi}$
 by repeated raising.
 
 The GH$^\nu$ operators can be expressed in terms of electron and vortex ladder operators by evaluating
 Eq. (\ref{eq:simplePH}) for the desired value of $m_\nu$.  The labels shown in the figure are complete in the ``bookend" cases $L, m_L=\pm L$,
 e.g.,  $[\boldsymbol{b}^{\dagger \, 2} \boldsymbol{\tilde{v}}^2]^2$ and $[\boldsymbol{\tilde{b}}^2 \boldsymbol{v}^{\dagger \, 2}]^2$.
  In other
 cases the lowering produces in general multiple terms, all of which have the same effect when acting on $|0_{N_{1 \over 2}} \rangle$.
  The labeling of the non-bookend cases is a shorthand, retaining one of several contributing terms that are generated
 under $\nu$-spin raising or lowering, but chosen to faithfully represent the underlying the mirror symmetry.
 
  We conclude that PH conjugation manifests itself in CF representations as an
 electron-vortex exchange symmetry, connecting mirror states around $m_\nu=0$: the mirroring operations are
$\boldsymbol{b}^\dagger \leftrightarrow 
 \boldsymbol{v}^\dagger$ and $\boldsymbol {\tilde{b}} \leftrightarrow \boldsymbol{\tilde{v}}$. 
 And although we use the ladder operators $\boldsymbol{\tilde{b}}$ and $\boldsymbol{\tilde{v}}$ to indicate the destruction of quanta in the
 half-filled state, alternatively we can make a standard canonical transformation, considering this instead to be creation of a hole,
 with the half-filled shell the particle and hole vacuum.   This would be an elegant way to formulate the vacuum state and
 its excitations.

 \begin{figure*}[ht]   
\centering
\includegraphics[width=0.75\textwidth]{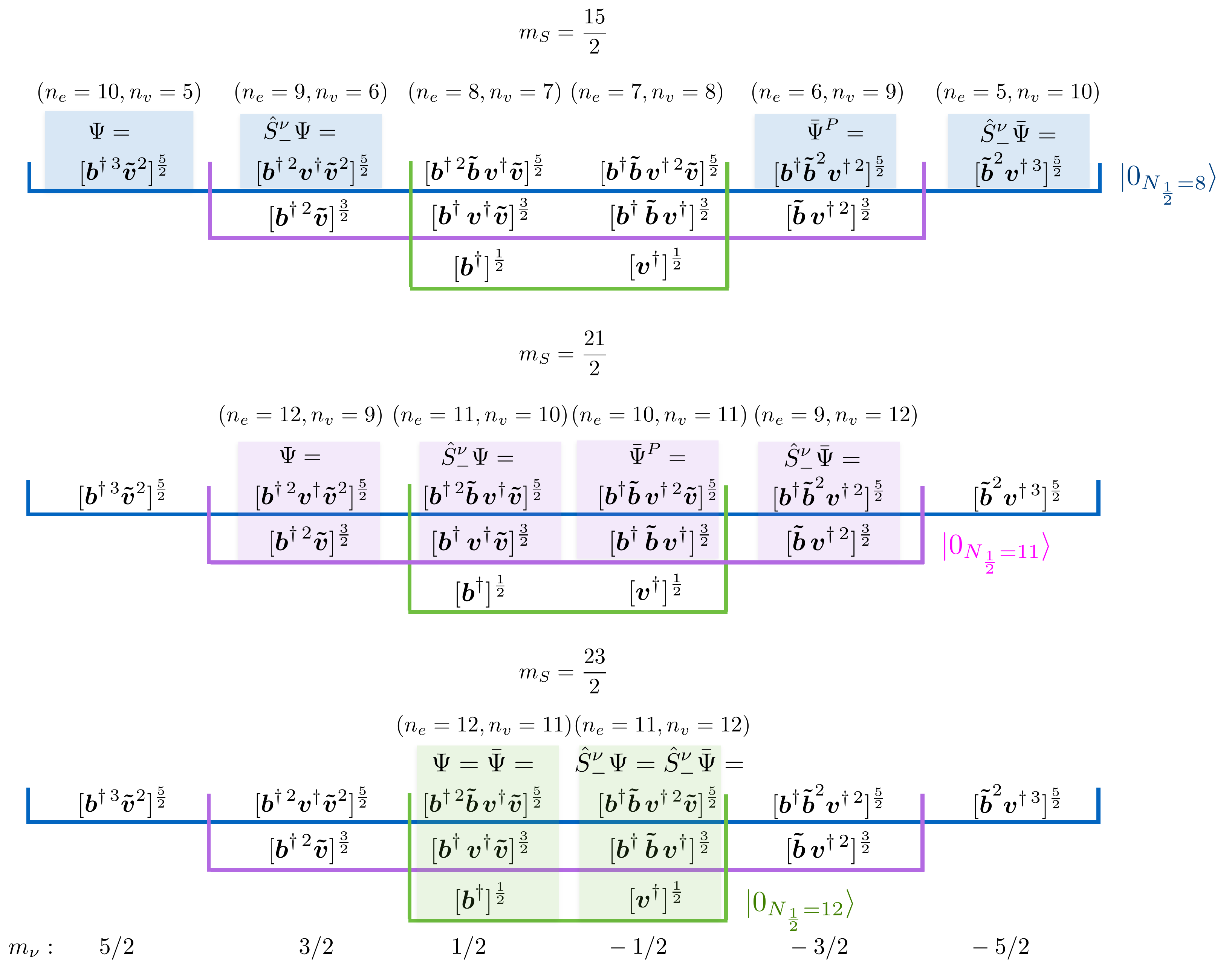}
\caption{As in Fig. \ref{fig:electronvortex_integral}, but for half-integer angular momentum.}
\label{fig:electronvortex}
\end{figure*}

 The half-integer angular momentum case is given in Fig. \ref{fig:electronvortex}.  
 (Contrast with the right panel of Fig. \ref{Fig:PHpattern}, the similar GH case where PH CFs belong to distinct multiplets.)  
 Here we must make a choice in describing the
 states nearest $m_\nu=0$, whether to treat them as particles or holes above $|0_{N_{1 \over 2}} \rangle$, which determines
 $N_{1 \over 2}$.  We choose to treat these states as single-particle excitations.  Then
 \begin{equation}
 N_{1 \over 2}=m_S+{1 \over 2}={N+\bar{N} \over 2}=
 p^2+2 L_{\mathcal{I}=1} p+L_{\mathcal{I}=1}-{1 \over 2}
 \label{eq:simpleVacuum}
 \end{equation}
 Otherwise the same electron-vortex mirror symmetry is found to describe the relations among PH conjugate CFs.
 
 The possibility of connections between PH symmetry and CF electron-vortex symmetry has been mentioned in previous
 work, including that of  Son \cite{Son1,Son2,Son3}, Geraedts et al. \cite{Geraedts},  and Metlitski and Vishwanath \cite{Metlitski}.  However we  know of no explicit demonstration of the relationship between these symmetries, other than that found here.   That
 may be because of the prevailing view that CF's are electrons coupled to two flux units, and thus not electron-vortex
 symmetric.

We have shown that $\nu$-spin creates the multiplets by which a CF evolves to its conjugate partner.   There is also interest in
PH evolution of many-CF wave functions, e.g., whether the $\nu<{1 \over 2}$ state with $p$ subshells  and the $\nu>{1 \over 2}$
 PH conjugate state with $p+1$ subshells could be built symmetrically from a vacuum state of filling $p+{1 \over 2}$, as discussed in \cite{Son1}.
 The successive building of the PH conjugate $\nu>{1 \over 2}$ state from from the $\nu<{1 \over 2}$ state can
 certainly be done, by following each $\nu$-spin lowering of all existing CFs, by the addition of a CF to the initially empty $\bar{\mathcal{I}}=1$
 subshell, $\bar{N}-N$ steps.
 It is attractive to do this, because the addition of a new zero-mode CF after every lowering $\hat{S}^\nu_-$ keeps the length
 of the vortex consistent with the number of CFs.  Thus every intermediate step produces a physical many-CF state.
 
 However, as the intermediate states correspond to a partially filled shell, these states are not unique, requiring some scheme
 to be defined.  The simplest procedure is to fill the single-CF states of the zero-mode subshell in the order descending $m_L$.
 Each many-CF state thus formed will have definite total $L^T$ and $m^T_L=L^T$, reaching a maximum at the half-filled shell,
 then declining after, with $L^T=m_L^T =0$ when the zero-mode subshell is filled.   
 
 For the case of half-integer angular momentum, $\bar{N}-N$ even, and there is a well-defined half-filled shell that, in our
 scheme, has maximum $L^T$ and $m_L^T=L^T$.  Starting from this state, one can reverse the order of operations described above, removing 
 a CF from the half-filled zero mode, then applying $\hat{S}^\nu_+$ to the remaining CFs; alternatively, from the same starting point,
 one can apply $\hat{S}^\nu_-$ and add a new CF.    These processes can be viewed as creating a hole or adding a particle to
 a vacuum state defined by the half-filled zero-mode shell, antialigned in angular momentum.  The $\bar{N}-N$ odd case
 differs only slighty.

\section{Dirac and Pauli Hamiltonians, \\
PH and Electron-Vortex Symmetry}
In the previous section we found that PH conjugation could be viewed as a mirror symmetry if we redefined the PH partners
to be those of Eq. (\ref{eq:partners}).  We now turn to the question of CF effective Hamiltonians, where these four
degrees of freedom also play an important role.

Below we  propose forms for the Pauli and
Dirac Hamiltonians that CFs satisfy.  We argue that the asymmetric subshell structure of PH conjugate pairs discussed above is the consequence of
an underlying isospectral Hamiltonian built out of the $\nu$-spin operators $\hat{S}^\nu_\pm$.  The construction identifies the
extra $\mathcal{I}=1$ subshell of the $\nu>{1 \over 2}$ state as the zero mode.  We show that the Hamiltonian can be linearized,
yielding a Dirac form, and note the similarities to Hamiltonians of the IQHE.
 
\subsection{Pauli and Dirac Hamiltonians for the IQHE}
Because of the IQHE-FQHE algebraic correspondence generated by $\hat{S} \leftrightarrow \hat{S}^\nu$, it is helpful to 
begin by reviewing the IQHE case.  The IQHE Pauli Hamiltonian \cite{Greiter} and solutions are
\begin{flalign}
&\hat{H}_P=
 {\hbar^2 \over 4 m_e  a_0^2} \,{1 \over m_S} \left( \hat{S}_- \hat{S}_+  +  \hat{S}_+ \hat{S}_- \right) \nonumber \\
&\hat{H}_P \mathcal{D}^L_{m_S \, m_L} =  {\hbar^2 \over 2 m_e  a_0^2} {1 \over m_S} (L(L+1)-m_S^2) \mathcal{D}^L_{m_S \, m_L}  \nonumber \\
&\equiv {p_P^2 \over 2m_e} \mathcal{D}^L_{m_S \, m_L} 
\label{eq:Pauli1}
\end{flalign}
For large $m_S>>\mathcal{I}$ the associated momentum scale is
\[ p_P \sim {\hbar \over a_0} \sqrt{2 \mathcal{I} -1}, ~~\mathcal{I}=1,2,3,... \]
as $L=m_S+\mathcal{I}-1$.

A related equation linear in momentum can be written
\begin{eqnarray} 
\hat{H}_D &=&  {\hbar  \over  R} c_B \left( \begin{array}{cc} 0 & \hat{S}_- \\ \hat{S}_+ & 0 \end{array} \right) = {\hbar  \over  a_0 \sqrt{m_S}} c_B \left( \begin{array}{cc} 0 & \hat{S}_- \\ \hat{S}_+ & 0 \end{array} \right) \nonumber \\
~  
\label{eq:Dirac}
\end{eqnarray}
where $c_B$ is a velocity.  This has the form of a massless Dirac equation.  This equation has been
studied previously \cite{Jellal, Arciniaga}.  Defining a momentum
\[ p_D = {\hbar \over a_0} \sqrt{2(\mathcal{I}-1) +{\mathcal{I} (\mathcal{I}-1) \over m_S}}  \rightarrow {\hbar \over a_0} \sqrt{2(\mathcal{I}-1)} \]
for $\mathcal{I}>1$ we find the two-component positive and negative energy solutions, $E=p_Dc_B$ and $E=-p_D c_B$, with 
eigenfunctions
\begin{equation}
 \left( \begin{array}{l} \mathcal{D}^L_{m_S \, m_L} \\ D^L_{m_S+1 \, m_L}  \end{array} \right)~~~~~~ \left( \begin{array}{l}~~ \mathcal{D}^L_{m_S \, m_L} \\ -D^L_{m_S+1 \, m_L}  \end{array} \right) 
 \end{equation}
respectively.  This solution has an electron of angular momentum $L$ residing equally
in two LLs, $\mathcal{I}$ and $\mathcal{I}-1$.  

The form of the IQHE Hamiltonians and the $\hat{S} \leftrightarrow \hat{S}^\nu$ correspondence suggests that one might
consider candidate CF Hamiltonians involving angular momentum and $\nu$-spin scalars such as $\hat{S}_+^\nu \hat{S}_-^\nu$.
A first question is whether such operators would capture the physics of CFs we have previously described.  One can show
\begin{eqnarray}
\hat{S}_+^\nu \hat{S}_-^\nu &=&-\boldsymbol{b}^\dagger \odot \boldsymbol{\tilde{b}} + \textstyle{1 \over 2} \boldsymbol{b}^\dagger \odot \boldsymbol{\tilde{b}} \, \boldsymbol{v}^\dagger \odot \boldsymbol{\tilde{v}} \nonumber \\
&&~~~~~~~~~~~~~~~~-[ \boldsymbol{b}^\dagger \otimes \boldsymbol{\tilde{b}} ]^1 \odot [ \boldsymbol{v}^\dagger \otimes \boldsymbol{\tilde{v}} ]^1 
\label{eq:isospectral}
\end{eqnarray}
The interaction terms involve scalar (antisymmetric) and rank-one (symmetric) pairs.   The former are number operators, while
the latter can be rewritten in terms of angular momentum operators.  One finds when acting on an arbitrary CF
\begin{flalign}
\hat{S}_+^\nu \hat{S}_-^\nu \rightarrow~~~~~~~~~~~~~~~~~~~~~~~~~~~~~~~~~~~~~~~~~~~~~~~~~~~~~~~~~~~  \nonumber \\
(N-1+2m_\nu) +\textstyle{1 \over 2} (N-1+2m_\nu)(N-1) -{2 \over \hbar^2} {\hat{L}}_e \odot {\hat{L}}_v 
\label{eq:hamS}
\end{flalign}
At fixed $m_\nu$ only the last term influences spectra.  This simple candidate Hamiltonian does incorporate the basic
correlation physics we have argued is important to CFs,
``measuring"  the opening angle between the 
electron and vortex spinors, and thus the number of favorable antisymmetric ($u(i) \cdot u(j)$) vs. unfavorable symmetric
($[u_(i) \times u(j)]^1$) pairs in the electron-vortex coupling.
The last term can be evaluated using
\[ -2 {\hat{L}}_e \odot {\hat{L}}_v = L(L+1) - L_e(L_e+1)-L_v(L_v+1) \]
so that when operating on a CF,
\begin{eqnarray}
\hat{S}_+^\nu \hat{S}_-^\nu &\rightarrow& (L+m_\nu)(L-m_\nu+1)
    \label{eq:Ls}
\end{eqnarray}
in agreement with results given previously.  We will see that this operator also has attractive properties in satisfying certain 
energy constraints imposed by PH symmetry.

%Recall that in our graphical depictions of PH conjugates states (Fig. \ref{Fig:SympPH}), there was 
%no PH partner for the IQHE.  The IQHE is the case where all CFs are in the zero mode,  annihilated by $A$.  
%Thus the above solution allows us to regard (correctly) the vacuum as the partner of this state, thus completing
%Fig. \ref{Fig:SymPH}.

\subsection{Connection to isospectral Hamiltonians}
Pauli Hamiltonians built on quadratic operator products of the form
\begin{eqnarray}
 \left( \begin{array}{cc}  A^\dagger A  & 0  \\ 0 & A A^\dagger   \end{array} \right) &\equiv& \left(\begin{array}{cc}  \bar{H} & 0 \\ 0 & H \end{array} \right) 
\end{eqnarray}
are isospectral, 
\begin{eqnarray}
\bar{H} |\bar{\Psi} \rangle = A^\dagger A |\bar{\Psi} \rangle = \bar{E} |\bar{\Psi} \rangle \Rightarrow ~~~~~~~~~~\nonumber \\
H A |\bar{\Psi} \rangle \equiv H |\Psi \rangle = A A^\dagger A |\bar{\Psi} \rangle = \bar{E} A |\bar{\Psi} \rangle \equiv E |\Psi \rangle
\end{eqnarray}
That is, if $|\bar{\Psi} \rangle$ is an eigenvector of $\bar{H}$, $A|\bar{\Psi} \rangle \sim |\Psi \rangle$ is an eigenvector of
$H$ with the same eigenvalue, provided 
$A |\bar{\Psi} \rangle$ exists.   In supersymmetric quantum mechanics, Hamiltonians of this form can account for partner spectra that are
identical apart from the presence of a zero mode, annihilated by $A$.   The
conventional normalization used in supersymmetric quantum mechanics to preserve the norm under conjugation is
\begin{equation}
|\Psi \rangle = {1 \over \sqrt{E}} A |\bar{\Psi} \rangle 
\end{equation}
Note that this identical to that of Eq. (\ref{eq:norm}), for $\hat{A}=\hat{S}^\nu_+$.  The use of Hamiltonians of the form
$\hat{S}^\nu_+ \hat{S}^\nu_-$ in the FQHE will produce isospectral CF eigenvalues when applied across multiplets.

%As the mapping to CFs produces a closed-shell wave function of the type encountered in noninteracting systems,
%we expect the underlying CF Hamiltonian to generate single-CF energies from which the many-body energy can be computed.
%An argument we previously gave in reference to the energies of $\nu$-spin multiplets -- integrating out deeply bound
%subshells -- could be applied to the $\bar{\nu} > \textstyle{1 \over 2}$ Laughlin subshell, removing $\bar{N}-N$ 
%CFs, and producing a reduced neutralizing positive background equivalent to $N$ positive charges.  Effectively this
%could be done by choosing the zero of energy to be that of the Laughlin $\bar{\nu}$ subshell CF energy, consistent
%with the identification of this energy with the Hamiltonian's zero mode.
%It is not implausible that the remaining $N$ CFs occupying the $p$ subshells of the $\nu$ and $\bar{\nu}$ states --
%CFs equivalent distributed over the sphere -- then would have equivalently spectra.   Though beyond the scope of this
%paper, it would be relatively easy to test this ansatz -- that CF spectra related by PH symmetry are isospectral -- by computing the single-CF energies, which can be done by
%removing a CF from the closed-subshell FQHE states.

\subsection{Correlations and the Pauli Hamiltonian}
While we have seen that CFs have the algebraic operator structure to support quadratic or linear Hamiltonians in $\nu$-spin,
the question remains whether this has any relationship to Hamiltonian
that would arise from an explicit treatment of the Coulomb interaction.  The approach we take below is based on an
evaluation of the two-particle correlation function, which we then relate to the subshell and operator structures discussed
in previous sections of this paper.

Interactions of electrons in a uniform neutralizing background $b$ consist of $b$-$b$, $e$-$b$, and $e$-$e$ terms, with
\[ V_{ee}= \alpha \hbar c ~{1 \over 2} \sum_{i \ne j=1}^N {1 \over |\boldsymbol{r}_i-\boldsymbol{r}_j|} \]
where the coordinate distance on the sphere is taken to be the chord.  It is convenient to expand this interaction
in multipoles, so that the monopole-monopole term can be employed to cancel the net $O(N^2)$ contribution from the sum of
$b$-$b$ and $e$-$b$ terms, leaving the residual interaction
\begin{equation}
 V_{ij} = {\alpha \hbar c \over R} {1 \over 2} \sum_{i \ne j=1}^N \left[ \sum_{\ell=1}^{2S} P_\ell(\cos{\beta_{ij}}) - {1 \over N-1} \right] 
\end{equation}
where $\beta_{ij}$ is the opening angle between the indicated electrons.  

The energy can be evaluated from
the two-particle correlation function \cite{Levesque}: utilizing the sphere's translational invariance and
homogeneity, one electron can be placed at $(\theta_1,\phi_1)=(0,0)$  and the second at $(\theta_2,\phi_2)=(\beta_{12},0) \equiv (\beta,0)$,
yielding
\begin{eqnarray}
g(\beta) &=& {8 \pi^2  \over \langle \Psi | \Psi \rangle } \int ~d\Omega_3 \cdots d\Omega_N~ \left| \Psi(0, \beta, \Omega_3, \cdots,
\Omega_N) \right|^2  \nonumber \\
1&=&\int g(\beta) \sin{\beta} d \beta   \nonumber \\
\langle V \rangle &=& \textstyle{N(N-1) \over 2} \int g(\beta) V_{12} (\beta) \sin{\beta} d\beta
\end{eqnarray}
where in the last line we equate the total energy of $N$ equivalent electrons to the pair energy times the number of pairs.

The simplest example is the case of $\nu=1$, where a straightforward calculation yields for arbitrary $N$
\begin{eqnarray}
 g^{\nu=1}_N(\beta) &=& \textstyle{N \over 2N-2} \sin^2{{\beta \over 2}} \left[1+\cos^2{{\beta \over 2}} + \cdots + [\cos^2{{\beta \over 2}}]^{N-2} \right] \nonumber \\
 &=& \textstyle{N \over 2N-2} \left[ 1- [\cos^2{{\beta \over 2}}]^{N-1} \right]
 \label{eq:nuis1}
 \end{eqnarray}
 The pair correlation function depends on the coordinate of the center $\beta/2$ and becomes uniform in the large $N$ limit apart
 from a vanishingly small region around $\beta/2=0$.  This can be seen to be correct physically by rewriting this answer in terms
 of the distance, using $\sin^2{\beta/2} = ({r \over 2R})^2 = {r^2 \over 2(N-1)a_0^2}$, then noting
 \begin{eqnarray}
  g^{\nu=1}(r) &\equiv& \lim_{N \to\ \infty} \left[ 1-\left[1-{r^2 \over 2(N-1) a_0^2} \right]^{N-1} \right] \nonumber \\
  &=& 1-e^{-{r^2 \over 2a_0^2}}, 
  \end{eqnarray}
 a familiar answer.  The energy can then be evaluated.  As the wave function has a noninteracting form, one is interested in the
 corresponding average energy per particle 
 \begin{eqnarray}
 {1 \over N} { \langle \Psi | V | \Psi \rangle \over \langle \Psi | \Psi \rangle} & \begin{array}{c} \\  \longrightarrow  \\ \scriptstyle{ N~ \mathrm{large}}   \end{array}&  - {\alpha \hbar c \over a_0} \sqrt{{\pi \over 8}} \left[ 1 + {5 \over 8N}+{57 \over 128N^2} \right] \nonumber \\
 &\equiv& \hbar \omega_{Coul}~ \epsilon_{\nu=1}^N
 \end{eqnarray}
 where $\hbar \omega_{Coul}$ is the Coulomb energy scale ${\alpha \hbar c \over a_0}$ and $\epsilon_{\nu=1}^N =   \sqrt{{\pi \over 8}} \left[ 1 + \cdots \right]$ is the dimensionless energy per particle for $\nu=1$, which becomes a constant for large $N$.
 The single-particle energy -- the energy of electron 1 in the field generated by all $N-1$ neighboring electrons -- would be twice this number.  
 
 Because PH conjugate states have the same CF subshell structure apart from the extra ``zero-mode" subshell of the $\nu>{1 \over 2}$ state,
 one might envision ``integrating out" the zero-mode shell, rendering the conjugate states algebraic identical, and perhaps
 relating their energies. Physically this makes some sense, as a filled subshell carries no angular momentum, so such an integration
 might add a uniform negative charge to the positive background charge, reducing the latter from a total of $\bar{N}$ charges to $N$ 
 charges, while not perturbing significantly the angular momentum of the subshells containing the remaining $N$ CFs.  
 Much more could be said about this idea, and its associated assumptions.  Unfortunately, if one sets to this task
 directly, several tricky issues arise.  However, it turns out can gain insight through a simpler procedure, calculating the 
 two-body correlation functions for conjugate states.   While we have not succeeded in evaluating the
 correlation functions analytically for arbitrary $N$ and $\nu$, analytic calculations are possible for small $N$.
 
 After this is done for various GH$^2$ CF PH conjugate states, we express the result as
   \begin{equation}
   g_{\bar{N}}^{\bar{\nu}} = a_\nu  \, {g}_{{N}}^{\nu} + a_1 \,  g_{N+\bar{N}}^{\nu=1}+ \,  g_\epsilon
  \label{eq:coreps}
  \end{equation}
  where we set
 \[ a_\nu \equiv {N(N-1) \over \bar{N}(\bar{N}-1)}~~~a_1 \equiv {\bar{N}(\bar{N}-1) -N(N-1) \over \bar{N}(\bar{N}-1)}, \]
 which defines $g_\epsilon$, and by our normalization condition requires
 \[ \int g_\epsilon(\beta) \sin{\beta} d\beta = 0 \]
 
 When such a calculation is done in cases where wave functions are unique and thus
 GH$^2$ PH symmetry is exact (e.g., $N$=3, $\bar{N}$=4, $S$=3) we find
 $g_\epsilon \equiv 0$.  In other small-$N$ cases we tested, the $g_\epsilon$ component of $g_{\bar{N}}^{\bar{\nu}}$ generates typically $\sim 0.05\%$ of the total many-electron energy.
 
 $g_\epsilon$ is a measure of the PH 
 symmetry breaking in the GH$^2$ construction, which employs distinct composite fermion forms for conjugate $\nu$ and
 $\bar{\nu}$ states, rather than computing the later from the former using PH symmetry.
 As total energies are obtained by summing the correlation energy over pairs, one can readily rewrite Eq. (\ref{eq:coreps}) 
 as (assuming exact PH symmetry)
 \begin{equation}
 \bar{N} \epsilon_{\bar{\nu}}^{\bar{N}} = N \epsilon_{\nu}^N + (\bar{N}-N) \epsilon_{\nu=1}^{\bar{N}+N}
 \end{equation}
 which one recognizes as equivalent to the PH relationship between total (many-electron) energies obtained by M\"{o}ller and Simon \cite{MS}.
 This result has added significance for the GH$^2$ construction because of its subshell structure, specifically the zero-mode
 Laughlin-like level that contains $\bar{N}-N$ CFs. The zero-mode CF energy thus must absorb the interactions
 among the $\bar{N}-N$ electrons associated with this subshell as well as with the uniform distribution of $N$ CFs that reside
 in the $p$ higher closed subshells.  We assign the $\bar{N}-N$ CFs in the lowest subshell of $\bar{\nu}> {1 \over 2}$ states
 the energy
\begin{equation}
H_\mathrm{zero ~mode} = \hbar \omega_{Coul} ~ \epsilon_{\nu=1}^{2S+1} 
\label{eq:zp}
\end{equation}
 
 %This decoupling, in our view, has important implications for how one should view the PH symmetry of FQHE
 %wave functions, when expressed in terms of CFs.
 %It is helpful to go back to Fig. \ref{Fig:PHfilledf}.
 %We see at $m_\nu={1 \over 2}$ a pattern of great symmetry: equal numbers of CF particle and hole subshells that are degenerate,
 %and thus easily associated with a Dirac PH picture.  We are arguing here that this picture is not limited to $m_\nu={1 \over 2}$,
 %but in fact extends to all of the figure's panels, once one recognizes the the CF mapping effectively removes the interactions
 %of $\bar{N}-N$ particles.   
 %Once this is done, we can treat $\nu <{1 \over 2}$ and $\nu>{1 \over 2}$ PH  state pairs 
 %as mirror $N$-particle CF states, occupying degenerate levels: there is no need to deal further with the $\bar{N}-N$ CFs in the
 %zero-mode subshell.  This observation relates to one of Son's arguments that struck us as awkward:
 %that the realization of PH symmetry might requires one to introduce a $(p+{1 \over 2})$-shell vacuum state.   This kind of picture
 %is forced on one by identifying CFs too strongly with electrons.  
 %In our view, the picture we have presented is more natural and elegant:  there are effectively only $p$ H and $p$ P
 %subshells -- carrying the same angular momentum quantum numbers, exactly degenerate, but labeled by distinct $m_\nu$ --
 %that are relevant to PH CF symmetry.
 
 The effective Hamiltonian describing the remaining $p$ valence subshells then must be symmetric around
 the PH pivot at $m_\nu={1 \over 2}$: this is clearly the case for  $\nu={1 \over 3} \leftrightarrow \bar{\nu}={2 \over 3}$, 
 as there is one such subshell; but it is also required for the towers of angular momentum subshells built successively on
 these, as there is no other physically reasonable way to preserve the PH energy relation.  Our limited 
 GH$^2$ CF algebra allows very few possibilities: the simplest operator with the necessary attributes is
 \[   {1 \over m_S} \hat{S}^\nu_+ \hat{S}^\nu_- \]
 The operator annihilates CFs in the zero-mode $\bar{\nu} > {1 \over 2}$ subshell, while producing identical eigenvalues,
 when acting on PH conjugate CFs (FQHE CFs in the same field $m_S$, with the same $L$, and $\bar{m}_\nu=1-m_\nu$). Thus it preserves the PH  energy constraint described above,
 subshell by subshell.    Would such an operator produce a physically reasonable CF spectrum?   
 For fillings of fixed $s$ but large $N$, and thus the sequence
 $\nu={1 \over 3}, \, {2 \over 5}, \, {3 \over 7}, ....$ (so $s=0, \, {1 \over 2}, \, 1, ...$), we find that the spread in subshell operator
 eigenvalues is ${4s \over 4s+3}$.  Thus the spread increases gently with $s$ to an asymptote of 1, a necessary condition to 
 keep subshell energies well
 defined in the large-$N$ limit.  As there are $2s+1$ subshells and thus $2s$ splittings, we see that the splitting between
 neighboring subshells (so $\Delta L=1$) is on average $2 \over 4 s +3$, narrowing as one approaches the half-filled
 shell.  The splittings are uniform at large $N$, consistent with the physics behind GH$^2$ CFs, with the number of broken
 antisymmetric pairs increases linearly as one ascends the angular momentum tower.  
 One can also check the large-$N$ half-filled case, where the evolution with $N=(2s+1)^2$ is along the trajectory $\ell=s$,
 instead of fixed $s$.
 The total spread among the eigenvalues at large $N$ is 1:  the number of subshells grows as $\sqrt{N}$, while the splitting between neighboring subshells goes as $1/\sqrt{N}$.
 The evolution from $s=0$ ($\nu={1 \over 3}$) to the half-filled shell is smooth and continuous, in all aspects.
 
 Now in principle we can add to this simple Hamiltonian other terms that are mirror symmetric around $m_\nu={1 \over 2}$,
 thus preserving the PH energy relation, e.g., 
 one could try a suitable combination of $S_0^{\nu \, 2}$ and $S_0^\nu$ (though other aspects of combining such operators
 are not attractive).  But the simplest choice is ${1 \over m_S} \hat{S}^\nu_+ \hat{S}^\nu_-$, in 
 combination with an additive constant.  As we are interested in the large-$N$ limit, we fit the two parameters in this Hamiltonian
 to the average energy per particle we compute for $\nu={1 \over 3}$, ${2 \over 5}$, and ${3 \over 7}$ extrapolated to large $N$.
 This extrapolation employs a spherical area density correction introduced by Morf et al. \cite{Morf}, which reduces the
 impact of finite-$N$ corrections on the extrapolation.  We find dimensionless average energies of -0.4098, -0.4326, and -0.4419,
 respectively, for these three fillings. 
 The single-CF Hamiltonian that emerges is
 \begin{eqnarray}
&&H_2^\mathrm{eff} = \hbar \omega_{Coul} \left[-0.635 + 0.339 {1 \over m_S} \hat{S}^\nu_+ \hat{S}^\nu_-\right] \nonumber \\
&&H_\mathrm{zero ~mode} = -\hbar \omega_{Coul} \sqrt{\pi \over 8} \sim \hbar \omega_{Coul} \left[ -0.627 \right]
\label{eq:zppp}
\end{eqnarray}
The subscript denotes that we employ two parameters.
From this Hamiltonian one would calculate $\epsilon_\nu$ for a given fractional filling by summing over the occupied
subshells, weighted by their occupancy, then dividing by the particle number.  The two parameters were fit to three $\epsilon_\nu$ values:
the average deviation is $\sim 0.03$\%, which likely is comparable to the extrapolation errors
we introduced in estimating the three $\epsilon_\nu$s.  

We have treated the zero-mode separately from the valence subshells: thus the two equations above.  But we
observe that the Hamiltonian's constant term -0.635 is within about 1\% of the zero-mode energy.  With this hint from the numerics, 
we repeat the fit, constraining the constant term in this way, yielding a one-parameter effective Hamiltonian that applies
equally to the zero-mode and to the $p$ conjugate subshells,
\begin{equation} 
H_1^\mathrm{eff} = \hbar \omega_{Coul} \left[-\sqrt{\pi \over 8}  + 0.3385 {1 \over m_S} \hat{S}^\nu_+ \hat{S}^\nu_-\right] 
\end{equation}
The fit remains very good, with deviations in predicted $\epsilon_\nu$ on average of less than 0.1\%.   Finally, as this Hamiltonian
has only one parameter, we need only fit one $\epsilon_\nu$, which we choose to be $\epsilon_{\nu={1 \over 3}}$.
That is, the coefficient 0.3385 must be a simple function of $\epsilon_{\nu={1 \over 3}}$.
We finally obtain an extremely simple effective Hamiltonian that respects PH symmetry
\begin{equation}
H_0^\mathrm{eff} = \hbar \omega_{Coul} \left[-\sqrt{\pi \over 8}  + {3 \over 2} \left(\epsilon_{\nu={1 \over 3}}+\sqrt{{\pi \over 8}} \right){1 \over m_S} \hat{S}^\nu_+ \hat{S}^\nu_-\right] 
\label{eq:Hsim}
\end{equation}
The average deviation in predicting $\epsilon_\nu$ for $\nu={2 \over 5}$, ${3 \over 7}$, ${4 \over 7}$, and ${3 \over 5}$ is $\sim 0.2$\%.
The GH$^2$ construction thus generates a high quality effective Hamiltonian that, given $\epsilon_{\nu={1 \over 3}}$ and
$\epsilon_{\nu=1} = -\sqrt{{\pi \over 8}}$, then provides a parameter-free description of all other fractional fillings.  The
operator appearing in Eq. (\ref{eq:Hsim}) can alternatively be expressed as
\[ \hat{S}_+^\nu \hat{S}_-^\nu =2 \hat{S}_0^\nu + \hat{S}_-^\nu \hat{S}_+^\nu= \hat{S}_0^\nu +{1 \over 2} \left( \hat{S}_+^\nu \hat{S}_-^\nu+\hat{S}_-^\nu \hat{S}_+^\nu \right). \]

\begin{figure*}[ht]   
\centering
\includegraphics[width=1.0\textwidth]{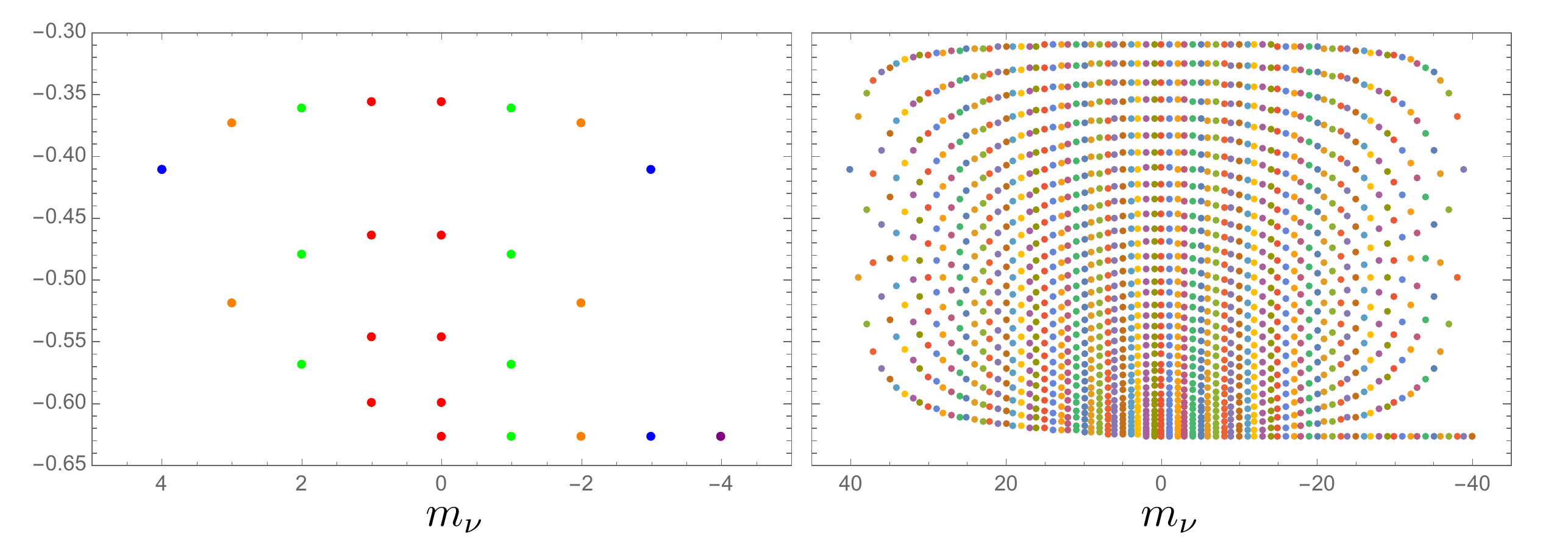}
\caption{The CF subshell energies from Eq. (\ref{eq:Hsim}) for the multiplets where $\ell+s=4$ (left) and $\ell+s=40$ (right), showing the
symmetry around $m_\nu={1 \over 2}$ that is broken by the appearance of the zero mode at $m_\nu=0$, continuing for negative
$m_\nu$ (more easily
seen in the left panel).  The level spacings evolve from linear to quadratic with decreasing $|m_\nu|$, as does the level density.
In the left panel, energy degeneracies (apart from the zero mode) are indicated by color. }
\label{fig:levels}
\end{figure*}  

This formula is valid for large $N$, as the fixed values for the two constants were fitted to extrapolated energies.
 Following our earlier discussion, we 
can now be more precise about CF subshell structure.  For fixed filling (fixed $p$)
and $\nu<{1 \over 2}$ the energy of the first ($\mathcal{I}=1$) subshell -- the Laughlin subshell with anti-aligned 
electron and vortex spinors -- is
\[ \epsilon_{\nu=1} + {3 \over 2p+1} ( \epsilon_{\nu={1 \over 3}} -\epsilon_{\nu={1}} )  \]
Thus for some large fixed $N$, with increasing density (decreasing $m_\nu$), $p$ increases, so the energy of the lowest subshell drops monotonically, reaching 
the asymptote $\epsilon_{\nu=1}$ at $\nu \sim {1 \over 2}$.   At this point the Laughlin subshell becomes the zero-mode subshell,
with fixed energy $\epsilon_{\nu=1}$ for all $m_\nu<0$.

For large $N$, as one moves from large $|m_\nu|$ toward $m_\nu=0$ (that is, $\nu={1 \over 2}$), the number of subshells increases,
while the gap between neighboring subshells decreases.  For example, the total spread between the highest and lowest subshells for $\nu<{1 \over 2}$, 
\[ {3 (p-1) \over 2p-1} ( \epsilon_{\nu={1 \over 3}} -\epsilon_{\nu=1} )  \rightarrow   {3 \over 2} ( \epsilon_{\nu={1 \over 3}} -\epsilon_{\nu=1} ), \]
increases monotonically, reaching the indicated asymptote near $\nu \sim{1 \over 2}$.  
The level spacing of the $p$ shells, for large but fixed $p$ but with $N \rightarrow \infty$ (assumptions that keep us away from
$m_\nu=0$), is uniform.  

In Fig. \ref{fig:levels} all of the CF subshell energies are plotted as a function of $m_\nu$,
for fixed $\ell+s=4$ and $40$.  Recall, from the column labels of Figs. \ref{Fig:Even} and \ref{Fig:Odd}, that each choice of ($\ell+s$,$m_\nu$)
corresponds to a unique FQHE state.  Thus the energies plotted are those of the CFs in those filled-subshell FQHE states.  The
mirror symmetry of energies around $m_\nu={1 \over 2}$ is the correspondence between CFs energies of PH states, for the $p$
subshells in common.

The figure shows the linear level spacing away from $m_\nu=0$ smoothly evolving into
a nonlinear $L^2$ pattern at $m_\nu=0$.   The spectrum remains approximately quadratic in the region around $m_\nu=0$, e.g., 
for large-$N$ trajectories of the type $\ell-s=m_\nu$=constant that we discussed  much earlier.
The patterns are mirror symmetric around $m_\nu={1 \over 2}$
except for the zero mode that appears at $m_\nu  =0$ and continues for $m_\nu$ negative. 

The average energy per particle can be calculated by summing over the occupied CF subshells, weighted by their occupancy, and 
dividing by $N$, yielding
\begin{eqnarray}
 \epsilon_\nu &=& \epsilon_{\nu=1} + {3 \over 2} (\epsilon_{\nu={1 \over 3}} -\epsilon_{\nu=1}) {2l (s+1)  \over (2l+1)(2s+1)+l-s-1}  \nonumber \\
 &=&  \epsilon_{\nu=1} + {3 \over 2} (\epsilon_{\nu={1 \over 3}} -\epsilon_{\nu=1}){ 2m_S-N+1  \over 2m_S} \nonumber \\
 &\rightarrow&    \epsilon_{\nu=1} + {3 \over 2} (\epsilon_{\nu={1 \over 3}} -\epsilon_{\nu=1}) (1-\nu)
 \end{eqnarray}
 for large $m_S$.  Thus average energies for FQHE states of filling ${1 \over 3} \le \nu \le 1$ are linear in $\nu$, which is in quite good accord with numerical calculations.
This simple result is obtained despite considerable structure in the subshell spectroscopy, apparent in
 Fig. \ref{fig:levels}.
 
  \subsection{Mirror symmetry and the linearized Hamiltonian}
  Although PH symmetry links states equidistant from
  $m_\nu={1 \over 2}$, and thus appears not to be a mirror symmetry around $m_\nu=0$, a version of 
  mirror symmetry reappears when the Pauli Hamiltonian is linearized, related also to electron-vortex symmetry.  We
  describe the linearization process here, then return to the question of a valence operator representation that exhibits
  these symmetries explicitly.
 
 We found above that PH-conjugate FQHE states satisfy the Schroedinger equations
\begin{eqnarray}
\left[ {\hbar \omega^\prime \over 2} +\Delta \hbar \omega_{Coul} \hat{S}^\nu_+ \hat{S}^\nu_-\right] \tilde{\Psi}^{N \, L}_{m_\nu \, m} &=&E \tilde{\Psi}^{N \, L}_{m_\nu \, m} \nonumber \\
\left[ {\hbar \omega^\prime \over 2} +\Delta \hbar \omega_{Coul} \hat{S}^\nu_+ \hat{S}^\nu_-\right] \tilde{\Psi}^{P \,\bar{N} \, L}_{1-m_\nu \, m} &=&E \tilde{\Psi}^{P \,\bar{N} \, L}_{1-m_\nu \, m} 
\end{eqnarray}
We have now included the magnetic energy, defining $\hbar \omega^\prime \equiv \hbar \omega +2 \epsilon_{\nu=1} \hbar \omega_{Coul}$ and
$\Delta \equiv {3 \over 2} (\epsilon_{\nu={1 \over 3}}-\epsilon_{\nu=1}){1 \over m_S} >0$.   The simplest interpretation of this equation is that
all CFs have the same mass ${\hbar \omega^\prime \over 2}$, which can be trivially removed by defining an interaction energy 
$E^\prime=E-{\hbar \omega^\prime \over 2}$.   CFs in the zero-mode contribute only through their mass.  We implicitly
remove the zero mode above by adding a superscript $P$ on $\tilde{\Psi}^{P \,\bar{N} \, L}_{1-m_\nu \, m}$, to indicate we 
are not  considering the zero-mode $m_\nu=L+1$ case.  This ensures that the first equation above always exists.  No information is
lost by doing this projection, as the zero mode is trivial. 

One can make use of the Pauli Hamiltonian's isospectral form,
multiplying on the left by $\hat{S}^\nu_-$, and noting
\[ \hat{S}^\nu_- \tilde{\Psi}^{N \, L}_{m_\nu \, m} \sim \tilde{\Psi}^{N+1 \, L}_{m_\nu-1 \, m}~~~~~~~\hat{S}^\nu_- \tilde{\Psi}^{P \,\bar{N} \, L}_{1-m_\nu \, m} \sim \tilde{\Psi}^{\bar{N}+1 \, L}_{-m_\nu \, m} \]
to obtain two additional equations
\begin{eqnarray}
\left[ \Delta \hbar \omega_{Coul} \hat{S}^\nu_- \hat{S}^\nu_+\right] \tilde{\Psi}^{N+1 \, L}_{m_\nu-1 \, m} &=&E^\prime \tilde{\Psi}^{N+1 \, L}_{m_\nu-1 \, m} \nonumber \\
\left[ \Delta \hbar \omega_{Coul} \hat{S}^\nu_- \hat{S}^\nu_+ \right] \tilde{\Psi}^{\bar{N}+1 \, L}_{-m_\nu \, m} &=&E^\prime \tilde{\Psi}^{\bar{N}+1 \, L}_{-m_\nu \, m} 
\end{eqnarray}

These results tell us that, in fact, CF PH symmetry is precisely a mirror-symmetric electron-vortex symmetry (see next subsection), relating states 
of opposite $m_\nu$
\begin{equation}
 \Psi \leftrightarrow \hat{S}^\nu_- \bar{\Psi}^P ~~~\mathrm{and}~~~\hat{S}^\nu_- \Psi \leftrightarrow \bar{\Psi}^P .
 \label{eq:pairing}
 \end{equation}
The relationships are illustrated for one case in Fig. \ref{fig:electronvortex}.

These expressions involve single CFs:  if one envisions lowering all the CFs in a FQHE wave function in this way, the electron or CF number will
clearly not change, even though the CF superscript $N$ or $\bar{N}$ is incremented.  This incrementing is thus a lengthening of the vortex only.  
The natural interpretation of the wave functions $\hat{S}^\nu_- \Psi$ and  $\hat{S}^\nu_- \bar{\Psi}^P$ is not unlike
the familiar construction of fractionally charged excitations of a FQHE state \cite{Laughlin},
\[ \Psi(z_1, \cdots, z_N)  \rightarrow \prod_{i=1}^N (z_i-z_0) ~ \Psi(z_1, \cdots,z_N)  \]
where a defect at $z_0$ is added.  In the present spherical case the application of $\hat{S}^\nu_-$ on each of $N$ CFs in a subshell
keeps fixed the number of quanta in each electron wave function, but adds a quantum labeled by $N+1$ to each vortex.  Consequently
the transformed wave function has the form
\[ \hat{S}^\nu_- \Psi(1,\cdots,N) \rightarrow \left[ \Psi^\prime(1,\cdots,N)\right]^{N \over 2}  \odot [u_{N+1}]^{N \over 2} \]
as the lowering operator is an angular momentum scalar.   Basically this expression represents the introduction of a defect not at
one point, but symmetrically distributed over the sphere due to the scalar product. This result helps one see the
qualitative physics behind the identification $\hat{S}^\nu_- \Psi \leftrightarrow \bar{\Psi}^P$.  Taking the $\nu={1 \over 3}, {2 \over 3}$ case
as an example, $\Psi$ would be the Laughlin subshell, with electron and vortex antialigned, the favored configuration;  
but $\bar{\Psi}^P$ is the $\mathcal{I}=2$ subshell of the $\bar{\nu}={2 \over 3}$ state, where the CFs have one broken scalar pair.  Thus it makes some physical sense that $\bar{\Psi}^P$
is conjugate not to $\Psi$ but to $\hat{S}^\nu_- \Psi$, as the introduction of a defect restricts the portion of the sphere occupied
by the $N$ electrons, necessarily breaking pairs.

The four-fold degeneracy (though two Hamiltonians are involved) leads to a familiar Pauli equation 
\begin{widetext}
\begin{equation}
 \left( \begin{array}{cccc} \hat{S}^\nu_+\hat{S}^\nu_- & 0 & 0 & 0 \\ [1.2ex]
0 &   \hat{S}^\nu_-\hat{S}^\nu_+ & 0 & 0 \\ [1.2ex]
0 & 0 &  \hat{S}^\nu_+\hat{S}^\nu_-  & 0 \\ [1.2ex]
0 & 0 & 0 &  \hat{S}^\nu_-\hat{S}^\nu_+ \end{array} \right) \left( \begin{array}{l} \tilde{\Psi}^{N \, L}_{m_\nu \, m} \\ [1.2ex]
 \tilde{\Psi}^{N+1 \, L}_{m_\nu-1 \, m} \\ [1.2ex] \tilde{\Psi}^{\bar{N} \, L}_{1-m_\nu \, m} \\ [1.2ex]
 \tilde{\Psi}^{\bar{N}+1 \, L}_{-m_\nu \, m} \end{array} \right) = {E^\prime \over   \Delta \,  \hbar \omega_{Coul}} \left( \begin{array}{l} \tilde{\Psi}^{N \, L}_{m_\nu \, m} \\ [1.2ex]
 \tilde{\Psi}^{N+1 \, L}_{m_\nu-1 \, m} \\ [1.2ex] \tilde{\Psi}^{\bar{N} \, L}_{1-m_\nu \, m} \\ [1.2ex]
 \tilde{\Psi}^{\bar{N}+1 \, L}_{-m_\nu \, m} \end{array} \right) 
 \end{equation} 
\end{widetext}
where $E^\prime = E-\hbar \omega^\prime/2$. 

This form is identical to that we found in our earlier discussion of the IQHE Pauli Hamiltonian (though in that case we
needed to ignore $\hat{S}_0$ to put the Hamiltonian in this form).  The only difference is replacement
$\hat{S} \rightarrow \hat{S}^\nu$.  As before we can take the square root of the Pauli
Hamiltonian to obtain a Dirac-like Hamiltonian, but now in $\nu$-spin space
\begin{equation}
\hat{H}_D \Psi  \sim \left( \begin{array}{cccc}  0 & \hat{S}^\nu_+ & 0 & 0\\ [1.2ex]  \hat{S}^\nu_- & 0 & 0 & 0 \\ [1.2ex] 0 & 0 & 0 & \hat{S}^\nu_+ \\ [1.2ex]
 0 & 0 &\hat{S}^\nu_- & 0  \end{array} \right)  \left( \begin{array}{l} \tilde{\Psi}^{N \, L}_{m_\nu \, m} \\ [1.2ex]
 \tilde{\Psi}^{N+1 \, L}_{m_\nu-1 \, m} \\ [1.2ex] \tilde{\Psi}^{\bar{N} \, L}_{1-m_\nu \, m} \\ [1.2ex]
 \tilde{\Psi}^{\bar{N}+1 \, L}_{-m_\nu \, m} \end{array} \right) 
\end{equation}
The matrix is two-by-two block diagonal, and thus has the same solutions given previously for the two-by-two IQHE
Dirac equation -- except LLs are now CFs FLL subshells, and states in neighboring magnetic fields now becomes states
of neighboring filling $\nu$ (or $m_\nu$).
The analogy with textbook four-component Dirac equations is rather close:  a $\nu$-spin doublet coupled by the
operators $\hat{S}^\nu_\pm$, paired with a similar doublet for the electron-vortex-symmetric PH conjugate partners.

While the proposed Hamiltonian assigns the same energies to conjugate subshells, 
the corresponding energies of many-CF wave functions differ because of the $\bar{N}-N$ CFs occupying the
extra zero-mode subshell of the $\bar{\nu} >{1 \over 2}$ FQHE state.  These CFs contribute to the energy through the
mass term 
that they carry in common with all of the other CFs.   There appears to be is an interesting weak-field limit
\[ \hbar \omega = 2 \sqrt{\pi \over 8} \hbar \omega_{Coul} \]
in which this mass vanishes.  It appears possible to select a field where the mass vanishes, the zero-mode energy
vanishes, and the PH conjugate many-electron states are degenerate.  Even more exotically, one could envision applying
a weak field on the plane with its strength smoothly varying over the plane.  This could be arranged to produce
domains of positive and negative mass, separate by a wall where that mass vanishes.  It would be interesting
to explore such possibilities further.

\section{Summary}
One of the important ideas in the FQHE, emerging from the work of Jain and others,
is that this open-shell, interacting problem might have a much simpler underlying CF representation.
To our knowledge the GH$^2$ construction is the only explicit demonstration of such a mapping,
providing simple analytic trial wave functions for small $N$ and all relevant $\nu$ that are in excellent agreement with those obtained
from exact diagonalizations of the Coulomb interaction.  The CFs that emerge from the construction are spherical
products of electron and vortex spinors, with the spinors formed from the aligned coupling of the underlying
ladder operators.  The CF mapping takes a complicated, interacting one-component electron system and maps it
into a much simpler, noninteracting, two-component system consisting of tightly coupled electrons and vortices.

The purpose of this paper has been to clarify the properties and symmetries of GH$^2$ CFs -- their quantum numbers, the algebraic
relationships that link states of different fillings, the connections between these symmetries and the underlying microscopic structure
of CFs, and the effective Hamiltonians that CFs satisfy.  The elegant
description of CFs derived here -- electron and vortex ladder-operator excitations of the half-filled shell -- helps one to see
these connections.

Crucial to this work is the recognition of the role of $\nu$-spin and its connections to electron-vortex symmetry.  $\nu$-spin does
not appear to have been considered previously.   Given the critical role that the analogous symmetry of isospin has played
in clarifying the spectroscopy and symmetries of QCD's CFs, protons and neutrons, this is surprising to us.

We summarize our main results:
\begin{enumerate}
\item We showed that the GH$^2$ wave functions can be written in  equivalent
CF and hierarchical forms, consisting respectively of 1) $p$ closed subshells occupied by $N$ electrons dressed by their
intrinsic wave functions,
or alternatively 2) $N/p$ vortices carrying angular momentum $L=p/2$ dressed by their own intrinsic wave functions, separated
on the sphere via spin-spin correlations. 
\item  We introduced the $\nu$-spin quantum number $m_\nu$, a second magnetic index, with
$N,L,m,m_\nu$ then identified as the complete set of CF quantum labels.  We described the algebra of the associated
operators $(\hat{S}_0^\nu, \hat{S}^\nu_\pm)$.  We also employed $\nu$-spin to redefine the ladder operators for the FQHE as
four-component objects, $\boldsymbol{d}^\dagger_{m_\nu \, m_L}$, thereby treating electron and vortex excitations symmetrically.
\item  We showed that all FQHE wave functions can be arrayed in mirror symmetric multiplets indexed by $m_\nu$, with the vortex
length $N-1$ held constant across multiplets.   The closed-shell $\nu$-spin multiplet states of maximum $|m_\nu|$ 
were identified as the incompressible FQHE states.  This connects states in mirror $\nu$-spin pairs:  such states have the same
$N$ and shell structure, but conjugate fillings ${p \over 2p+1}$ and ${p \over 2p-1}$, depending on the sign of $m_\nu$.
\item  We then described these states more elegantly as aligned valence ladder operators acting on a scalar half-filled intrinsic
state. This identifies electron particle-hole conjugation $\boldsymbol{b}^\dagger \leftrightarrow \boldsymbol{\tilde{b}}$ as the
transformation that, if performed on each CF, converts  an $N$-particle FQHE state of filling $p \over 2p+1$ into its 
$\nu$-spin mirror of filling $\nu={p \over 2p-1}$.
\item  We then turned to PH conjugation: in the context of our constant $N$ multiplets, PH conjugate states belong to different multiplets,
with $\bar{m}_\nu = 1-m_\nu$, and with fillings ${p \over 2p+1}$ and ${\bar{p} \over 2 \bar{p} -1}={p+1 \over 2p+1}$.
\item  We showed that the $\nu$-spin raising operator $[\hat{S}^\nu_+]^{\bar{N}-N}$ transforms the $\nu>{1 \over 2}$ state into
the PH conjugate state with $\nu<{1 \over 2}$, annihilating the $\bar{N}-N$ CFs in the $\bar{\mathcal{I}}=1$ subshell in the process.  
We observed that the magnetic field strength $m_S$, not $N$, is preserved under $\nu$-spin. This led to a second set of CF multiplets in
which $m_S$ is constant, and to an associated valence representation of the CFs as operators GH$^\nu$
acting on the half-filled subshell.  In these new multiplets,
CFs associated with the $\nu<{1 \over 2}$ state, $\mathcal{I}=1,\dots,p$, and those associated with the $\nu>{1 \over 2}$ state,
$\bar{\mathcal{I}}=2,\dots,\bar{p}=p+1$, are members of the same $\nu$-spin multiplet.
\item  This allowed us to show that PH symmetry is manifested in CF representations of wave functions as a
 electron-vortex exchange operation, $\boldsymbol{b}^\dagger \leftrightarrow \boldsymbol{v}^\dagger$.
 We also found that PH conjugation is a mirror symmetry, though one that links the conjugate states
 \[ \Psi \leftrightarrow \hat{S}^\nu_- \bar{\Psi} ~~~\mathrm{and}~~~\hat{S}^\nu_- \Psi \leftrightarrow \bar{\Psi}^P \]
\item  From a combination of physics arguments and numerical explorations, we concluded that the effective CF Hamiltonian
is isospectral, formed from $\hat{S}^\nu_+ \hat{S}^\nu_-$.  The $\bar{\mathcal{I}}=1$ subshell containing $\bar{N}-N$ 
CFs was identified as the zero mode; CFs in this subshell have a mass in common with all others CFs, but no other
contribution to their energies.  We showed that the Pauli
Hamiltonian can be linearized, cast into a Dirac form.   The degrees of freedom in the Dirac equation belong to the same
constant-$m_S$ multiplet and come in mirror pairs:  $\Psi \leftrightarrow \hat{S}^\nu_- \bar{\Psi}$ and $\hat{S}^\nu_- \Psi 
\leftrightarrow \bar{\Psi}^P$.
\end{enumerate}

Although we have worked on the sphere, our spherical results
carry over immediately to the plane:  the algebraic procedure for mapping spherical results
to the plane, described in \cite{GH2}, is based on the correspondence
between the spherical operators $L_x,L_y$ and the planar operators $p_x,p_y$.   This leads to appropriate analogs of the spherical scalar and tensor products - a rather elegant way to create multi-electron wave functions that behave
simply under translations, including scalars that are translationally invariant.

We thank John Wood for helpful discussions.   This material is based upon work 
supported in part through the US Department of Energy, Office of Science, by the
Office of Nuclear Physics under Awards DE-SC00046548 (Berkeley, WH) and
DE-AC02-05CH11231 (LBNL, WH),  and by  the Office of Basic Energy Sciences, Materials Sciences 
and Engineering Division
under Contract DE-AC02-05-CH11231 (LBNL, BK)  through the Scientific Discovery through 
Advanced Computing (SciDAC) program (KC23DAC Topological and Correlated Matter via Tensor 
Networks and Quantum Monte Carlo).   Addition support was provided by the
Peder Sather Center (DH, WH) and the Simons Foundation (WH).  Parts of the work was performed during visits to
the KITP (UC Santa Barbara), the INT (U. Washington), and CASS (UC San Diego), who WH thanks for their hospitality.


\begin{thebibliography}{000}

\bibitem{Laughlin} R. B. Laughlin, Phys. Rev. B{\bf 23}, 5632 (1981); B. I. Halperin, Phys. Rev. B{\bf 25}, 2185 (1982).

\bibitem{Tsui} D. C. Tsui, H. L. St{\"o}rmer, and A. C. Gossard, Phys. Rev. Lett. {\bf 48}, 1559 (1982);
H. L. St{\"o}rmer, Rev. Mod. Phys. {\bf 71}, 875 (199).

\bibitem{Jain} J. K. Jain, Phys. Rev. Lett. {\bf 63}, 199 (1989); J. K. Jain, S. A. Kivelson, and N. Trivedi,
Phys. Rev. Lett. {\bf 64}, 1297 (1990).

\bibitem{GH} J. N. Ginocchio and W. C. Haxton, Phys. Rev. Lett. {\bf 77}, 1568 (1996).

\bibitem{JainCF} J. K. Jain, Ann. Rev. Cond. Matt. Phys. {\bf 6}, 39 (2015).

\bibitem{dyakonov} M. I. Dyakonov, in Recent Trends in Theory of Physical Phenomena in High Magnetic Fields, ed. I. D. Vagner et al. (Kluwer
Academic Publishers, 2003), pp. 75-88 (arXiv:cond-mat/0209206).  We stress we are referring to the entire wave function expressed
as a single determinant, thus explicitly defining the CFs, not the operators that act on the half-filled shell, as the GH operators
were given as single determinants.

\bibitem{GH2} W. C. Haxton and D. J. Haxton, Phys. Rev. B {\bf 93}, 155138 (2016).

\bibitem{Note} One often hears the statement that the effective magnetic field itself vanishes at the half-filled shell.  
This reflects the confusion between
$\hat{S}^\nu$ and $\hat{S}$ addressed in this paper.  The quantity vanishing is $m_\nu$, the electron-phonon number 
difference, not the applied field.  The behavior of correlated electrons with $m_\nu$ is analogous to the behavior of uncorrelated
IQHE electrons with magnetic field strength $S$, but the driving physics is clearly distinct, as the IQHE behavior persists if the 
Coulomb interaction is turned off.

\bibitem{Haldane} F. D. M. Haldane, Phys. Rev. Lett. {\bf 51}, 605 (1983).

\bibitem{Greiter} Martin Greiter, Phys. Rev. B {\bf 83}, 115129 (2011).

\bibitem{Varshalovich} D. A. Varshalovich, A. N. Moskalev, and V. K. Kersonskii, ``Quantum Theory of Angular Momentum,"
World Scientific (Singapore) 1988.

\bibitem{Arciniaga} M. Arciniaga and M. R. Peterson, Phys. Rev. B {\bf 94}, 035105 (2016).

\bibitem{Jellal} Ahmed Jellal, Nucl. Phys. B {\bf 804}, 361 (2008).

\bibitem{Halperin} B. I. Halperin, Phys. Rev. Lett. {\bf 52}, 1583 (1984).

\bibitem{deb1} J. K.Jain, Indian J. of Phys. {\bf 88}, 915 (2014).

\bibitem{deb2} P. Sitko, K.-S. Yi, and J. J. Quinn, Phys. Rev. B {\bf 56}, 12417 (1997).

\bibitem{deb3} T. H. Hansson, M. Hermanns, and S. Viefers, Phys. Rev. B {\bf 80}, 165330 (2009); J. Suorsa, S. Viefers, and
T. H. Hansson, Phys. Rev. B {\bf 83}, 235130 (2011); T. H. Hansson, M. Hermanns, S. H. Simon, and S. F. Viefers, 
Rev. Mod. Phys. {\bf 89}, 25005 (2017).

\bibitem{deb4} P. Bonderson, Phys. Rev. Lett. {\bf 108},066806 (2012).

\bibitem{deb5} Z. Papic, Phys. Rev. B {\bf 87}, 245315 (2013).

\bibitem{GirvinLN} S. M. Girvin, The Quantum Hall Effect: Novel Excitations and Broken Symmetries, in ``Topological Aspects of Low-Dimensional Systems.," ed.  A. Comtet, T. Jolicoeur, S. Ouvry , and F. David, vol 69 (Springer, Berlin, 2002) pp. 53-175.

\bibitem{Laugh2} R. B. Laughlin, Chapter 3 in "The Quantum Hall Effect," ed. R. E. Prange and S. M. Girvin (Springer-Verlag, New York, 1990).

\bibitem{Weinberg} S. Weinberg, Phys. Lett. B {\bf 251}, 288 (1990); Nucl. Phys. B {\bf 363}, 3 (1991); Phys. Lett. B {\bf 295}, 114 (1992).

\bibitem{Son1} D. T. Son, talk presented at the Nambu Memorial Symposium (March, 2016) 
\url{http://nambu-symposium.uchicago.edu/talks/Son_nambu2016.pdf}

\bibitem{Son2} D. T. Son, Prog. Theor. Exp. Phys. {\bf 2016}, 12C103 (2016).

\bibitem{Son3} M. Levin and D. T. Son, Phys. Rev. B {\bf 95}, 125120 (2017).

\bibitem{Geraedts} S. D. Geraedts {\it et al.}, Science {\bf 352}, 197 (2016).

\bibitem{Metlitski} M. A. Metlitski and A. Vishwanath, Phys. Rev. B {\bf 93}, 245151 (2016).

\bibitem{Levesque} D.Levesque, J. J. Weis, and A. H. MacDonald, Phys. Rev. B{\bf 30}, 1056 (1984).

\bibitem{MS} G. M\"{o}ller and S. H. Simon, Phys. Rev. B {\bf 72}, 045344 (2005).

\bibitem{Morf} R.Morf and B. I. Halperin, Z. Physik B {\bf 68}, 391 (1987).


\end{thebibliography}
\end{document}